\documentclass{aa}
\usepackage{graphicx}
\usepackage{wasysym}
\usepackage{amsmath}
\usepackage{multirow}
\def\arxivprefixe{arXiv}
\def\arxivprefixesep{:}

\usepackage[separate-uncertainty]{siunitx}
\DeclareSIUnit{\mearth}{M_\oplus}
\DeclareSIUnit{\mjup}{M_{Jup}}
\DeclareSIUnit{\rearth}{R_\oplus}
\DeclareSIUnit{\rjup}{R_\jupiter}
\DeclareSIUnit{\rsol}{R_{\odot}}
\DeclareSIUnit{\msol}{M_{\odot}}
\DeclareSIUnit{\au}{au}
\DeclareSIUnit{\year}{yr}
\DeclareSIUnit\erg{erg}

\usepackage{tikz}
\usetikzlibrary{positioning}

\usepackage{txfonts}
\usepackage{subfigure}
\usepackage{color}

\usepackage{hyperref}
\hypersetup{pdfauthor=Remo Burn, pdftitle=Toward a Population Synthesis of Disks and Planets. I., breaklinks=true, colorlinks=true, urlcolor=blue, linkcolor=blue, citecolor=blue, bookmarksopen=true}

\renewcommand*\P[1]{Paper \uppercase\expandafter{\romannumeral #1\relax}}

\defcitealias{Sellek2020a}{S20}
\def\ctS20{\citetalias{Sellek2020a}}
\def\cpS20{\citepalias{Sellek2020a}}

\renewcommand*\d[0]{\mathrm{d}}

\begin{document} 
	
   \title{Toward a Population Synthesis of Disks and Planets}
   \subtitle{I. Evolution of Dust with Entrainment in Winds and Radiation Pressure}
   \titlerunning{Disk Population Synthesis I: Dust with Entrainment and Radiation Pressure}

   \author{R. Burn \inst{\ref{mpia}}, A. Emsenhuber \inst{\ref{lmu}}, J. Weder \inst{\ref{unibe}}, O. V\"olkel \inst{\ref{mpia}}, H. Klahr \inst{\ref{mpia}}, T. Birnstiel\inst{\ref{lmu},\ref{orig}}, B. Ercolano\inst{\ref{lmu},\ref{orig}}, C. Mordasini\inst{\ref{unibe}}}
   \authorrunning{R.~Burn et al.}

   \institute{Max-Planck-Institut f\"ur Astronomie, K\"onigstuhl 17, 69117 Heidelberg, Germany,\label{mpia}  
        \email{burn@mpia.de}
         \and
         University Observatory, Faculty of Physics, Ludwig-Maximilians-Universit\"at M\"unchen, Scheinerstr. 1, 81679 Munich, Germany \label{lmu}
         \and
         Physikalisches Institut, University of Bern, Gesellschaftsstrasse 6, 3012 Bern, Switzerland \label{unibe}
         \and
         Exzellenzcluster ORIGINS, Boltzmannstr. 2, D-85748 Garching, Germany \label{orig}
    }

   \date{accepted for publication in Astronomy \& Astrophysics on July 9, 2022}
 
  \abstract
   {Millimeter astronomy provides valuable information on the birthplaces of planetary systems. In order to compare theoretical models with observations, the dust component has to be carefully calculated.}
   {Here, we aim to study the effects of dust entrainment in photoevaporative winds and the ejection and drag of dust due to effects caused by radiation from the central star.}
   {We improved and extended the existing implementation of a two-population dust and pebble description in the global Bern/Heidelberg planet formation and evolution model. Modern prescriptions for photoevaporative winds were used and we account for settling and advection of dust when calculating entrainment rates. In order to prepare for future population studies with varying conditions, we explore a wide range of disk-, photoevaporation-, and dust-parameters.}
   {If dust can grow to pebble sizes, that is, if they are resistant to fragmentation or turbulence is weak, drift dominates and the entrained mass is small but larger than under the assumption of no vertical advection of grains with the gas flow. For the case of fragile dust shattering at velocities of \SI{1}{\meter\per\second} -- as indicated in laboratory experiments --, an order of magnitude more dust is entrained which becomes the main dust removal process. Radiation pressure effects disperse massive, dusty disks on timescales of a few 100 Myr.}
   {These results highlight the importance of dust entrainment in winds as a solid mass removal process. Furthermore, this model extension lies the basis for future statistical studies of planet formation in their birth environment.}

   \maketitle
\section{Introduction}
\label{sec:introduction}

The recent advances in millimeter astronomy, mainly thanks to ALMA, enable the detailed study of dust content in protoplanetary disks surrounding young stars, where planets are forming. Surveys of star-forming regions \citep[e.g.][]{Ansdell2016,Ansdell2017,Ansdell2018,Pascucci2016,Barenfeld2016} give a statistical overview of the properties of these disks. This data provides unprecedented evidence on planet formation processes. To make use of these constraints, it is required to model the dust evolution in theoretical studies of planet formation. This has been addressed by a number of dedicated studies in different numbers of dimensions and degrees of complexity \citep[e.g.][for a review]{Birnstiel2016}. Simplified models of the process \citep[e.g. two-population model by][]{Birnstiel2012} benefit from fast computational times and therefore usability in planetary population synthesis studies \citep{Voelkel2020}. However, these models were initially developed to predict pebble fluxes for constraining planetary growth without the comparison to millimeter observations in mind.

In particular, more research is needed for cases and disk regions where the bulk of the solids do not grow to pebble size and do not drift toward the star \citep{Weidenschilling1977} and additionally, where the gas is not viscously accreted toward the star. This happens commonly if the solids are fragile \citep[as indicated in recent laboratory experiments by][]{Steinpilz2019} or if turbulence is vigorous.

To that end, it is important to include additional processes influencing the dust and pebbles in these regions. An example process is dust entrainment in photoevaporative winds \citep{Owen2011a,Facchini2016,Hutchison2016,Franz2020,Franz2022,Franz2022a}. The research interest in this topic lay mainly on the potential observability of entrained dust in winds until \citet{Sellek2020a}, hereafter \ctS20, accounted for dust entrainment in an evolutionary model for the first time. They find that a significant mass fraction of dust can be removed by this process in disks subject to external photoevaporation. Similarly, dust entrainment in X-ray photoevaporative winds was included in the dust evolution calculations of \citet{Garate2021a} without in-depth analysis and discussion of this particular effect.

Other effects with possible influence on the dust mass content are known from the fields of debris disk and Solar System studies. We will study here the effects caused by radiation pressure which leads to Poynting-Robertson drag or direct ejection of dust \citep{Robertson1937,Burns1979,Klahr2001,Wyatt2008}. The requirement for these processes to become efficient is an optically thin disk for radiation to reach the dust grains.

Here, we extend the two-population dust evolution model of \citet{Birnstiel2012} implemented in a global model of planet formation and evolution \citep{Emsenhuber2020a,Voelkel2020} to include these processes. This marks an important step toward a planetary population synthesis model which at the same time can be tested against millimeter observations. Furthermore, we assess the effect of dust entrainment in photoevaporative winds using the modern prescriptions for internal XEUV photoevaporation by \citet{Picogna2019} -- recently updated by \citet{Picogna2021} and \citet{Ercolano2021} -- and for the external one by \citet{Haworth2018}. At the same time, we further explore the effects of settling of dust grains \citep[similar to][]{Franz2022} counteracted by vertical transport \citep{Booth2021}.

This paper is the first part of a series that includes the training of a statistical, surrogate model for fast comparison to observations in part two, and a Bayesian retrieval of initial disk conditions for unperturbed disks based on observations in part three. We present a complete gas and dust disk model description in Sect. \ref{sec:model} before the effects of different model choices (Sect. \ref{sec:model_comparison}) and parameters (Sect. \ref{sec:disk_param}) are shown. We put the results into perspective in Sect. \ref{sec:discussion} before summarizing the findings and concluding (Sect. \ref{sec:conclusion}).

\section{Model}
\label{sec:model}
\subsection{Viscous gas disk}
\label{sec:gas_disk}
Owing to the fact that this work is mostly about disks, we recall here the general description of a standard (see also \citealp{Emsenhuber2020a}), one-dimensional, viscous disk. Initially, the gas disk surface density profile as a function of distance to the star $r$ follows
\begin{equation}
\label{eq:initial_surf_dens}
\Sigma_\mathrm{g}(t=0) = \Sigma_{\mathrm{g},0} \left(\frac{r}{r_{0}}\right)^{-\beta_\mathrm{g}}\exp{\left(-\left(\frac{r}{r_\mathrm{out}}\right)^{2-\beta_\mathrm{g}}\right)}\left(1-\sqrt{\frac{r_\mathrm{in}}{r}}\right)\,,
\end{equation}
where $\beta_\mathrm{g}$ is the slope of the gas disk, $r_{0} = \SI{5.2}{\au}$ is chosen as a reference distance, $r_\mathrm{out}$ is the outer exponential cut-off radius, and $r_\mathrm{in}$ is the disk inner edge.

The evolution over time $t$ of the surface density $\Sigma_{\rm g}=\rho_g H_{\rm g} \sqrt{2\pi}$ for the midplane volume density $\rho_g$ of a viscous disk with Keplerian orbital velocities $\Omega_\mathrm{K} = \sqrt{G M_\star/r^3}$ as dictated by mass and angular momentum conservation is \citep{Weizsacker1948,Lynden-Bell1974,Pringle1981}
\begin{equation} 
\label{eq:disk_evol}
\frac{\d \Sigma_{\rm g}}{\d t} = \frac{3}{r}\frac{\d }{\d r} \left( r^{1/2} \frac{\d}{\d r}\left(r^{1/2} \nu\Sigma_{\rm g}\right)\right)-\dot{\Sigma}_\mathrm{int}-\dot{\Sigma}_\mathrm{ext}\,,
\end{equation}
where the viscosity $\nu = \alpha c_{s} H_{\rm g}$ is parameterized using the \citet{Shakura1973} $\alpha$-viscosity, $c_{s} = \sqrt{k_{\rm B} T_\mathrm{mid}/(\mu\, m_\mathrm{H})}$ is the midplane isothermal sound speed, $H_{\rm g} = c_{s}/\Omega_\mathrm{K}$ is the local disk scale height, $k_{\rm B}$ Boltzmann's constant, $\mu$ the mean molecular weight in units of hydrogen atom masses $m_\mathrm{H}$, and $\dot{\Sigma}_\mathrm{int}$ and $\dot{\Sigma}_\mathrm{ext}$ are additional terms for the losses by internal and external photoevaporation, respectively (see below).

The disk midplane temperature $T_\mathrm{mid}$ is given following the analytical approximation from \citet{Nakamoto1994}
\begin{equation}
\label{eq:Tmid}
T_\text{mid}^4 = \frac{1}{4\sigma}\left(\frac{3\kappa_R \Sigma_{\rm g}}{4} + \frac{1}{\kappa_P \Sigma_{\rm g}} \right) \dot{E}_\text{visc} + T_\text{surf}^4\,,
\end{equation}
where $\sigma$ is the Stefan-Boltzmann constant and the viscous energy dissipation rate is
\begin{equation}
\dot{E}_\mathrm{visc} = \Sigma_{\rm g} \nu \left( r \frac{\d \Omega}{\d r} \right)^2\ = \frac{9}{4} \Sigma_{\rm g} \nu \Omega_K^2
\end{equation} for a Keplerian disk.
For the ambient temperature due to stellar irradiation, we approximate

\begin{equation}
T_\mathrm{surf}^4 \approx T_\star^4 \left[\frac{2}{3\pi}\left(\frac{R_\star}{r}\right)^3+\frac{1}{2}\left(\frac{R_\star}{r}\right)^2\frac{H_{\rm g}}{r}\left(\frac{\d \ln H_{\rm g}}{\d \ln r} -1\right)\right] + T_\mathrm{irr}^4 + T_\mathrm{env}^4\,,
\label{eqn:tl_term}
\end{equation}

where we used $\d \ln H_{\rm g} / \d \ln r = 9/7$ following \citet{Chiang1997} as in \cite{Hueso2005}. Furthermore, we used a background temperature $T_\mathrm{env} = \SI{10}{\kelvin}$ and a direct irradiation term stemming from radially illuminating the disk through the midplane $T_\mathrm{irr}^4 = L_\star/(16 \pi r^2\sigma) e^{-\tau_\mathrm{mid}}$ with the radial optical depth through the midplane $\tau_\mathrm{mid} = \int \rho_\mathrm{g,mid}\kappa_R(\rho_\mathrm{g,mid},T_\mathrm{mid})\d r$.
Here and in Eq. \eqref{eq:Tmid}, $\kappa_P$ is the Planck mean opacity and $\kappa_R$ is Rosseland's mean opacity. Values for those opacities are used following \citet{Bell1994} including interstellar dust and molecular contributions. In future works, we will couple the dust evolution module (Sect. \ref{ssec:dust_evolution}) to the temperature structure. For this project, however, there is no feedback from those sub-modules onto the disk temperature or gas scale height.

\subsection{Photoevaporation}
The influence of high-energy radiation driving photoevaporation of the gas disk was taken into account with two prescriptions in this work. 

\subsubsection{Internal}

The internal photoevaporation caused by the central star follows the prescription of \citet{Picogna2019}, \citet{Ercolano2021}, and \citet{Picogna2021}. For this purpose, we need the stellar X-ray luminosity $L_\mathrm{X}$. The total mass loss rate as function of stellar luminosity is computed following \citet{Ercolano2021} and as function of stellar mass following \citet{Picogna2021}. The mass loss profiles were obtained from \citet{Picogna2019,Picogna2021}.

To compute the total mass loss rate as a function of a given stellar mass and X-ray luminosity $L_\mathrm{X}$, we proceeded as follows. First, we determined the part of the luminosity that is in the soft X-ray band (\num{0.1} to \SI{1}{\kilo\electronvolt}), which controls the mass loss rate \citep{Ercolano2021}. For this purpose, we performed a regression on the results of \citet{Ercolano2021}, which gives
\begin{equation}
    L_\mathrm{X,soft}/L_\mathrm{X} = 0.47333 - 0.055 \log_{10}{\left(L_\mathrm{X}/\SI{e30}{\erg\per\second}\right)}.
\end{equation}
The corresponding mass loss rate mass star can be obtained using Eq.~(5) of the same work,
\begin{equation}
\log_{10}{\left(\frac{\dot{M}_\mathrm{W,L}}{\si{\msol\per\year}}\right)}=a_\mathrm{S}\exp{\left(\frac{\left(\ln{\left(\log_{10}{\left(L_\mathrm{X,soft}/\si{\erg\per\second}\right)}\right)} - b_\mathrm{S}\right)^2}{c_\mathrm{S}}\right)} + d_\mathrm{S},
\end{equation}
where $a_\mathrm{S}=\num{-1.947e17}$, $b_\mathrm{S}=\num{-1.572e-4}$, $c_\mathrm{S}=\num{-2.866e-1}$, and $d_\mathrm{S}=\num{-6.694}$.

For the effect of the stellar mass on the mass loss rate, we used the results of \citet{Picogna2021} and their Eq.~(5), which gives
\begin{equation}
\dot{M}_\mathrm{W,M}=\SI{3.93e-8}{M_\odot\per\year}\frac{M_\star}{\SI{1}{M_\odot}}.
\end{equation}
To combine that with the stellar X-ray luminosity, we determined the equivalent X-ray luminosity for a \SI{1}{M_\odot}. For this we used the scaling of \citet{Gudel2007} so that
\begin{equation}
    \log_{10}\left(L_\mathrm{X,norm}/L_\mathrm{X}\right) = - 1.54\log_{10}\left(M_\star/\SI{1}{M_\odot}\right).
\end{equation}
The total mass loss rate for our stellar mass is then
\begin{equation}
\dot{M}_\mathrm{W}=\dot{M}_\mathrm{W,M}(M_\star)\frac{\dot{M}_\mathrm{W,L}(L_\mathrm{X,norm})}{\dot{M}_\mathrm{W,L}(\SI{e30.31}{\erg\per\second})},
\end{equation}
where \SI{e30.31}{\erg\per\second} is the mean stellar X-ray luminosity of a \SI{1}{M_\odot} star following \citet{Gudel2007}.

For the conversion into the surface density loss $\dot{\Sigma}_\mathrm{int}$, we used the fits to the same hydrodynamical simulations. All profiles have the same form,
\begin{equation}
\begin{split}
\dot{\Sigma}_\mathrm{int} &\propto 10^{a_X r_{10}^6+b_X r_{10}^5+c_X r_{10}^4+d_X r_{10}^3+e_X r_{10}^2+f_X r_{10}+g_X} \left(\frac{r}{\SI{1}{\au}}\right)^{-2} \\
& \quad \times  \left(6 a_X r_{10}^5+5 b_X r_{10}^4+4 c_X r_{10}^3 + 3 d_X r_{10}^2+ 2 e_X r_{10}+ f_X \right)\,,
\end{split}
\end{equation}
where $r_{10}=\log_{10}(r/\SI{1}{au})$.

This profile was normalized such that the integral $\int_0^\infty 2\pi r \dot{\Sigma}_\mathrm{int} \d r$ corresponds to the total mass loss rate $\dot{M}_\mathrm{W}$. The coefficients for a \SI{0.7}{M_\odot} star are given in \citet{Picogna2019}, while those for \num{0.1}, \num{0.3}, \num{0.5}, and \SI{1.0}{M_\odot} stars are in \citet{Picogna2021}. In case of stellar masses that were not modeled, we did not perform interpolation. Rather, we chose the closest stellar mass that was modeled and round toward lower stellar masses at mid distances, such that the profile of a \SI{0.2}{M_\odot} star mass is given by the parameter derived from the \SI{0.1}{M_\odot} case, for instance. We note that for this work we restricted our analysis to \SI{1}{\msol} stars but for the sake of completeness, we give the stellar mass dependency.

After an inner hole in the disk has opened an additional transition disk evaporation is employed, yielding higher total mass loss rates during this stage, where $\dot{M}_\mathrm{W}$ is no longer reached due to already depleted regions of the disk. The total transitional mass loss rate is given by
\begin{equation}
\log_{10}\left(\frac{\dot{M}_\mathrm{W,t}}{\SI{1}{\msol/yr}}\right) = 0.965 \times \log_{10}\left(\frac{\dot{M}_\mathrm{W}}{\SI{1}{\msol/yr}}\right) - 9.592\times 10^{-3} \times \frac{r_h}{(\SI{1}{AU})}
\end{equation}
and the profile is given by \citep{Picogna2019}
\begin{equation}
    \dot{\Sigma}_\mathrm{int,t} \propto a_t  b_t^x x^{(c_t-1)} \frac{x \ln(b_t) + c_t}{2 \pi \left(\frac{r}{\SI{1}{au}}\right)}
\end{equation}
where $a_t=0.11843$, $b_t=0.99695$, $c_t=0.14454$, and $x=(r-r_h)/\SI{1}{au}$ with $r_h$ being the hole radius, i.e. the radius where the integrated column density through the midplane reaches the maximum X-ray penetration depth of \SI{e22}{cm^{-2}} \citep{Owen2011,Owen2012}. A normalization factor is determined at the hole opening time to achieve a total mass loss rate of $\dot{M}_\mathrm{W,t}$. The employed criterion for hole opening is $r > \SI{2.35}{au} \left(M_{\star}/\SI{1}{\msol}\right) $ and, to prevent nonphysical evaporation, we restrict $r_h$ to be smaller than $\SI{120}{AU} \times \left(\frac{M_\star}{\SI{0.7}{\msol}}\right)$.

\subsubsection{External}
Using only internal X-ray and EUV evaporation yields \textit{relic disks}, i.e. rings of gas outside of $\sim \SI{100}{AU}$ that are not efficiently removed \citep{Owen2011,Owen2012}. Such remnants are not observed but can be efficiently removed when considering external photoevaporation through far-ultraviolet (FUV) irradiation.

Incorporating a model for external photoevaporation that is based on the FRIED grid (\textbf{F}UV \textbf{R}adiation \textbf{I}nduced \textbf{E}vaporation of \textbf{D}iscs) from \cite{Haworth2018} enables us to model the external photoevaporation as a function of the local FUV field strength $\mathcal{F}_\mathrm{FUV}$\footnote{$\mathcal{F}_\mathrm{FUV}$ is usually given in terms of $G_0$, which corresponds to the typical interstellar radiation field for FUV \citep{Habing1968}.}. For the implementation, we followed \citet{Weder2022}. They use linear interpolation to retrieve mass loss rates for given disk sizes and masses.

The outer radius of the disk has to be defined as where the disk transitions from optically thick to optically thin and can be located by searching for the maximum mass loss rate predicted from the FRIED grid for the outer disk region \citep[see discussion in][]{Sellek2020a}. 2D calculations from \cite{Haworth2019} showed that the mass loss rate is set entirely by the outer half of the disk and for the biggest part originates from the outer 10\% of the disc. Therefore, mass is considered to be removed uniformly from the outer 10\% of the disk ($\beta_\mathrm{ext}=0.9$), with $R_\mathrm{edge}$ being the outer edge of the disk and $\dot{M}_\mathrm{ext}$ being the corresponding mass loss rate retrieved from the FRIED grid.
\begin{equation}
    \dot{\Sigma}_\mathrm{ext} = 
    \begin{cases}
    0 & \qquad \mathrm{for} \quad r< \beta_\mathrm{ext} R_\mathrm{edge}\\
    \frac{\dot{M}_\mathrm{ext}}{\pi(R_\mathrm{edge}^2 - \beta_\mathrm{ext}^2R_\mathrm{edge}^2)}& \qquad \mathrm{for} \quad r \geq \beta_\mathrm{ext} R_\mathrm{edge}
    \end{cases}
\end{equation}
Note that in order to avoid numerical problems, a smoothing was applied to the transition at $\beta_\mathrm{ext}R_\mathrm{edge}$.

\subsection{Dust evolution}
\label{ssec:dust_evolution}
We used the two-population model of dust evolution described in \citet{Birnstiel2012} to get surface densities of the monomer grains and of larger-sized pebbles. This approach is an approximation to the full size distribution of grains and pebbles in the disk. The lower size $a_0$ represents the gas-coupled grains and the larger size $a_1$ the drifting pebbles. Due to the fact that there would be a distribution of sizes, the typical size $a_1$ is smaller than the maximum pebble size at a given location in the disk. For fragmentation limited regions, they are smaller by a factor  $f_f = 0.37$ and the typical size is given by
\begin{equation}
	a_1 = f_f \frac{2}{3\pi} \frac{\Sigma_{\rm g}}{\rho_{\rm s} \alpha_t} \frac{v_\mathrm{frag}^2}{c_s^2}\,,
	\label{eq:frag-size}
\end{equation}
where $c_s$ is the isothermal sound speed, $\rho_{\rm s} = \SI{1}{\gram\per\centi\meter\cubed}$ is the chosen grain density, and the vertical turbulence parameter $\alpha_t = \alpha$ controls the settling of dust grains 
\begin{equation}
H_{\rm d} \simeq H_{\rm g}\sqrt{\frac{\alpha_t}{\mathrm{St}}}
\label{eq:dust_scale_height}
\end{equation}
\citep{Youdin2007} as well as the relative velocities between grains \citep{Ormel2007,Birnstiel2016}. $\alpha_t$ was set here to the same value as $\alpha$ which drives the gas evolution of the disk. The last term, $v_\mathrm{frag}$ is the collisional velocity at which icy grains fragment \citep[e.g.][]{Blum2010}. It was previously estimated to lie at \SI{10}{\meter/s}, but could also lie an order of magnitude lower at \SI{1}{\meter\per\second} as recently indicated by laboratory experiments \citep{Gundlach2018,Musiolik2019,Steinpilz2019}.

In the case of growth being limited by drift, the large grain size is given by
\begin{equation}
a_1 = f_d \frac{2 \Sigma_{\rm dust} v_\mathrm{K}^2}{\pi \rho_{\rm s} c_s^2} \bigg|\frac{\d \ln P}{\d \ln r}\bigg|^{-1}\,,
\label{eq:drift-size}
\end{equation}
where a second free parameter $f_d = 0.55$ is fit to model outcomes \citep{Birnstiel2012}.

\citet{Birnstiel2012} further introduce a third parameter $f_m$ to relate the total surface density of all dust $\Sigma_{\rm dust}$ to the surface densities of the two size bins as
\begin{align}
\Sigma_1(r) &= \Sigma_{\rm dust}(r) f_m(r)\\
\Sigma_0(r) &= \Sigma_{\rm dust}(r) (1-f_m(r))\,,
\end{align}
where $f_m$ is set to 0.97 or 0.75 in the drift or fragmentation limited case, respectively.

The surface density of the two size bins is evolved as a combined advection/diffusion equation
\begin{equation}
\frac{\d \Sigma_{\rm dust}}{\d t} + \frac{1}{r} \frac{\partial}{\partial t} \left[ r \left( \Sigma_{\rm dust} \bar{u} - D \frac{\partial}{\partial r}\left(\frac{\Sigma_{\rm dust}}{\Sigma_{\rm g}}\right) \Sigma_{\rm g}\right)\right] = L\,,
\end{equation}
where $L$ can combine source or sink terms, $D = \alpha c_s^2 /\Omega_\mathrm{K}$ is the diffusion coefficient of the gas also used for particles here \citep[as justified for small $\mathrm{St}$, see][]{Birnstiel2012} and the mass weighted velocity $\bar{u}$ is given by 
\begin{equation}
\label{eq:ubar_comb}
\bar{u} = (1-f_m(r))u_0 + f_m(r) u_1\,.
\end{equation}
$\bar{u}$ combines the velocities of the large and small grain population with their velocities $u_0$ and $u_1$ given as the combined radial drift (azimuthal drag) and radial gas motion effects \citep{Whipple1972,Weidenschilling1977,Nakagawa1986,Garate2020}.

In order to increase the precision of our prescription in regions with high dust to gas ratios, we include the feedback of the dust onto the gas velocity, thus
\begin{equation}
\label{eq:reduced_drift}
u_\mathrm{g, red} = A u_{\rm g} - 2 B \eta v_\mathrm{K}\,,
\end{equation}
where we used the common definition of $\eta = - \frac{r}{2v_\mathrm{K}^2 \rho_{\rm g,mid}} \frac{\d P}{\d r}$. For simplicity, this reduced gas velocity is solely used for the dust evolution. Furthermore, we calculate the feedback coefficients assuming a mass-weighted, single dust size $\bar{a} = (1-f_m(r))a_0 + f_m(r) a_1$, thus Eqs. 29 and 30 from \citet{Garate2020} can be used:
\begin{align}
A &= \frac{\bar{\rho}_{\rm dg} + 1 + \bar{\mathrm{St}}^2}{(\bar{\rho}_{\rm dg}+1)^2 + \bar{\mathrm{St}}^2}\\
B &= \frac{\bar{\rho}_{\rm dg} \bar{\mathrm{St}}}{(\bar{\rho}_{\rm dg}+1)^2 + \bar{\mathrm{St}}^2}\,.
\end{align}

Here, and throughout this work, the Stokes number is calculated assuming the Epstein regime $\mathrm{St} = \frac{\pi a \rho_{\rm s}}{2 \Sigma_{\rm g}}$, where $a$ is either a mean or a distinct size of the two size bins. With the reduced radial gas velocity from Eq. \eqref{eq:reduced_drift}, the drift velocity of the two dust sizes used in Eq. \eqref{eq:ubar_comb} can be calculated following \citet{Nakagawa1986}
\begin{equation}
u_{1,2} = \frac{u_\mathrm{g,red}}{1+\mathrm{St}_{1,2}^2} - \frac{2\eta v_\mathrm{K}}{\mathrm{St}_{1,2} + (\mathrm{St}_{1,2} \varrho^2)^{-1}}\,.
\label{eq:radial_drift_w_feedback}
\end{equation}

Differing from the original \citet{Birnstiel2012} model, we include the factor $\varrho = \frac{\rho_{\rm g,mid}}{\rho_{\rm g,mid} + \rho_{\rm d,mid}}$ following \citet{Nakagawa1986} that reduces radial drift if dust midplane densities $\rho_{\rm d,mid}$ become comparable to the gas midplane density $\rho_{\rm g, mid}$. For this consideration, the dust midplane density is calculated assuming a Gaussian profile with a scale height of $H_{\rm d}$ (Eq. \ref{eq:dust_scale_height}) and a single mass-averaged size $\bar{a}$.

\subsection{Dust disk clearing by entrainment in photoevaporative winds}
\label{sec:late_stage}
This main model for the dust evolution was extended to account for a more realistic treatment in regions where the gaseous disk disperses. There, entrainment of dust grains in photoevaporative winds and radiation pressure on the grains (Sect. \ref{sec:radiation_pressure_model}) can become relevant. 
A variation to the model assuming settled dust with no vertical dust flux, in the following called the 'settling' model, is described in Appendix \ref{app:settled_entrain}.

Wherever photoevaporative winds launch gas from the disk surface, dust particles can be entrained in the wind \citep{Facchini2016,Hutchison2016,Franz2020}. Given a local gas surface density removal rate of $\dot{\Sigma}_{\rm g}$, the drag force in the Epstein regime on a spherical dust particle is
\begin{equation}
F_D = \frac{4\pi}{3} a^2 v_{\rm th} \rho_{\rm g} v_r = \frac{4\pi}{3} a^2 v_{\rm th} \frac{\dot{\Sigma}_{\rm g}}{\mathcal{F}}\,,
\end{equation}
where $v_{\rm th}$ is the thermal velocity of the gas in this wind and we followed \ctS20 in describing the radial velocity of the wind $v_r = \dot{\Sigma}_{\rm g}/(\mathcal{F}\rho_{\rm g})$ with a geometric factor $\mathcal{F}=\frac{H_{\rm g}}{\sqrt{r^2+H_{\rm g}^2}}$. The temperatures used to calculate the thermal velocity of the winds are \SI{1000}{\kelvin} \citep{Matsuyama2003} for external and \SI{2000}{\kelvin} for X-ray photoevaporation \citep{Picogna2019}. For a stationary grain, we can equate this drag force to the gravitational pull of the star to derive a critical size of
\begin{equation}
a < a_{\mathrm{ent,ext}} = \frac{v_{\rm th} \dot{\Sigma}_{\rm g} R^2}{\rho_{\rm s} \mathcal{F} G M_\star}\,.
\label{eq:aent_ext}
\end{equation}
This size limit is applicable for a non-moving particle and thus for externally launched winds.
For a given size distribution of the dust $n(a) \propto a^{q}$, where we assume $q=-3.5$ \citep{Dohnanyi1969,Birnstiel2016} and a size-independent dust bulk density $\rho_{\rm s}$, it is therefore possible to calculate the mass fraction of entrained particles \footnote{\ctS20 set $a_{\rm min} \approx 0$ in Eq. \eqref{eq:entrained_fraction}, while we keep the minimum value to have a higher precision in regimes where little growth occurred.}
\begin{equation}
f_{\rm ent} = \frac{a_{\rm ent}^{4+q} - a_{\rm min}^{4+q} }{a_{\rm max}^{4+q} - a_{\rm min}^{4+q}}\,.
\label{eq:entrained_fraction}
\end{equation}
The local change in dust surface density is then given by
\begin{equation}
\dot{\Sigma}_\mathrm{ent} = f_{\rm ent} \delta_{\rm dg} \dot{\Sigma}_{\rm g}\,,
\label{eq:dust_photoevap}
\end{equation}
where $\delta_{\rm dg}$ is the dust to gas ratio.

For internal photoevaporation, \citet{Booth2021} derived a more constraining limit taking into account vertical advection of dust grains from the midplane to the base of the flow. In this scenario, the strongest limitation on the maximum dust size that can be entrained originates from whether a grain can be lifted to the the region just below the ionization front which marks the base layer of an evaporative flow driven by XEUV radiation. We include the following advection-related limit whenever we consider internal photoevaporation:
\begin{equation}
a_{\mathrm{ent,int}} = \sqrt{\frac{8}{\pi}} \frac{\dot{\Sigma}_{\mathrm{int}}}{\rho_{\rm s} \Omega_\mathrm{K}} \frac{H_{\rm base}}{z_{\rm base}} \left( 1 + \frac{z_{\rm base}^2}{r^2} \right)^{3/2}\,.
\end{equation}
Here, we approximated $H_{\rm base} \approx H_{\rm g}$ which holds for a vertically isothermal disk. Furthermore, we used $z_{\rm base}/H_{\rm base} = 4$ as suggested by \citet{Booth2021}.

Overall, our implementation of dust entrainment follows \ctS20 for external photoevaporation and the suggested prescription of \citet{Booth2021} for internal photoevaporation. Lastly, it is not physically possible to entrain dust where no gas is left. Therefore, we check that the gas evaporation rate can not exceed the amount of local gas surface density multiplied by the timestep. This is in fact not always strictly true in evolving numerical models with finite sized time steps. Given a cell in which little gas was left, gas could diffuse into it during one timestep which would in principle allow for more entrainment which is now suppressed.

\subsection{Radiation pressure effects}
\label{sec:radiation_pressure_model}
Typically once the gas disk clears, remaining dust grains are exposed to direct irradiation from the star. The radiation pressure force acting on a spherical grain with surface area $A$ assuming perfect absorption (neglecting in particular Mie scattering) in radial direction is
\begin{equation}
F_{\rm rad} = \frac{L_\star A}{4\pi r^2 c}\,,
\end{equation}
where $L_\star$ is the stellar luminosity and $c$ is the speed of light in vacuum. Due to the identical scaling with $1/r^2$, the force can be related to the gravitational force by defining \citep[e.g.][]{Burns1979,Grun1985}
\begin{equation}
\beta = F_{\rm rad}/F_G = \frac{3 L_\star}{16 \pi G M_\star c \rho_{\rm s} a}\,,
\end{equation}
for a spherical particle with radius $a$ and density $\rho_{\rm s}$. This can then be used to modify the gravitational potential of the star $U' = - \frac{GM_\star (1-\beta)}{r}$. Escape of a particle from the system occurs if the total energy $E/m = v^2/2 + U' > 0$. For the case of a particle that was on a circular, Keplerian orbit ($v=v_\mathrm{K}$) before being irradiated, this leads to the condition $\beta > \frac{1}{2}$ \citep[e.g.][and references therein]{Burns1979,Klahr2001}. Therefore, particles below a critical size of
\begin{equation}
\label{eq:acrit_rad}
a_{\rm crit} = \frac{3 L_\star}{8\pi GM_\star c \rho_{\rm s}}
\end{equation}
are ejected from the system once radiation reaches them. Analogous to Eq. \eqref{eq:entrained_fraction}, a fraction of the local dust can be removed. For radiation pressure, which acts mainly after the gas disk has dissipated, it is appropriate to assume $q=-3.5$, the collision dominated size slope \citep[e.g.][]{Dohnanyi1969}. Modifications to this classical result would be more likely to arise in the presence of gas where particles drift \citep{Birnstiel2011} and not in the gas-free, debris disk environment.

The aforementioned fraction of dust is removed on either an orbital timescale or during one collisional timescales in the Brownian motion regime \citep{Birnstiel2016}
\begin{equation}
\label{eq:tcol}
t_{\rm col} = \frac{\pi a \rho_s}{6 \rho_{\rm d}}\sqrt{\frac{a^3 \rho_s}{3 k_{\rm B} T}}\,,
\end{equation}
where $\rho_{\rm d}$ is the volume mass density of dust distributed in the disk in contrast to the bulk density of the dust particles $\rho_{\rm s}$.
We used this timescale as the duration it takes to re-establish a collisional cascade size distribution after removing small particles for the case of $a_1 > a_{\rm crit}$ wherever it is longer than an orbital timescale.

To account for shielding of the disk, we calculated the optical depth $\tau_{\rm mid, dust}$ along a straight light ray through the midplane using the amount of dust in the small size bin given by the two-population model\footnote{$\tau_{\rm mid,dust}$ should in principle be equivalent to $\tau_{\rm mid}$ that is used for the temperature calculation (Sect. \ref{sec:gas_disk}). However, we refrained from consistently coupling the dust evolution to the temperature yet.}. The opacity of this micrometer dust follows \citet{Bell1994} and we assumed that the dust disk has a scale height of 0.04 for this calculation \citep[following the estimation of][for debris disks]{Thebault2009}. The amount of material that is removed is then reduced by the factor
\begin{equation}
    f_{\rm red} = \exp(-\tau_{\rm mid, dust}) + f_{\rm sur}\,.
\end{equation} This includes some radiation that can reach a razor-thin disk from the surface due to the geometrical extent of the star $f_{\rm sur} = 2 R_{\star}^3 / (3 \pi r^3)$. This way of accounting for shielding assumes that the photons that reach the dust still carry their full energy but not all the dust is affected immediately. We note that during the gas disk stage most of the disk is shielded from radiation through the midplane. Nevertheless, the innermost cells can be influenced and we allowed for radiation related effects also during the gas disk stage using the assumptions noted here (constant disk scale height, fragmentation dominated size distribution).

In addition to the possibility of reaching hyperbolic orbits due to radiation pressure, Poynting-Robertson drag can act \citep{Robertson1937}. In the rest frame of the sun, this effect can be seen as caused by the wavelength difference of emitted radiation by a moving particle \citep{Burns1979}. This leads to a decrease of eccentricities and semi-major axes of irradiated bodies with radii $a>a_{\rm crit}$.

Considering that eccentricities were damped in the gaseous disk very quickly and can not be efficiently increased by radiation pressure for large particles, we set the eccentricity to zero and the rate of increase of distance to the star as a result of Poynting-Robertson drag for circular orbits is then given by \citep[see e.g.][for derivations ]{Wyatt1950,Burns1979,Kobayashi2009}
\begin{equation}
\label{eq:PoyntingRobertson}
u_{r,\mathrm{rad}}  = - \frac{3 L_\star f_{\rm red}}{8 \pi c^2 a \rho_{\rm s} r}\,,
\end{equation}
which is added to $u_0$ and $u_1$ in Eq. \eqref{eq:ubar_comb}. We note that we reduced the velocity by $f_{\rm red}$ to account for optical shielding which makes the process inefficient in optically thick regions.

This simple approach to model Poynting-Robertson drag and radiation pressure neglects both stellar winds as well as the wavelength dependency of the absorption. For particles of comparable size to the typical wavelength of the radiation $\sim$\SI{0.1}{\micro\meter}, Mie scattering would occur. For those particles, collisions with radially outward moving stellar wind particles however becomes important which also leads to ejection \citep{Burns1979}.

In the absence of gas, the collisional growth prescription also needs to be adapted. As Brownian motion can be assumed to dominate the relative velocities, the collisional timescale assuming a size $a$ for all grains is again given by Eq. \eqref{eq:tcol}.  Assuming that the mass doubles each time two particles meet, this growth timescale leads to a size increase
\begin{equation}
\dot{a} = \frac{2\rho_{\rm d}}{\pi \rho_{\rm s}} \sqrt{ \frac{3 k_{\rm B} T}{\rho_{\rm s}} } a^{-3/2}\,
\end{equation}
which we use to evolve the larger representative size $a_1$ assuming here no radial mixing or drift after the gas has locally disappeared (see Sect. \ref{sec:caveats_2pop}).

\subsection{Model set-up}
\label{sec:setup}
\newcommand\T{\rule{0pt}{2.6ex}}       
\newcommand\B{\rule[-1.2ex]{0pt}{0pt}} 
\begin{table}
	\caption{Disk Parameters}
	\label{tab:modelParams}
	\centering
	\begin{tabular}{r c c}
		\hline\noalign{\smallskip} 
		Parameter & Symbol &  Value (variation)\\
		\hline\hline\noalign{\smallskip}  
		Stellar Mass & $M_{\star}$ & \SI{1}{\msol}\\
		Disk Mass & $M_{\rm disk}$ & \SI{0.012}{\msol} ($\pm$1 dex) \\
		Disk Viscosity & $\alpha $ & \SI{3.16e-3}{} ($\pm$1 dex)\\
		X-ray luminosity & $L_X$ & $\SI{1e29}{erg/s}$ ($\pm$1 dex)\\
		Gas slope & $\beta_\mathrm{g}$ & 0.95 (0.7,1.2)\\
		Inner edge & $r_{\rm in}$ & \SI{0.13}{\au}\\
		Exponential cut-off & $r_{\rm out}$ & \SI{30}{\au} (\SI{10}{\au}, \SI{100}{\au})\\
		Ext. UV field & $\mathcal{F}_{\rm FUV}$ & \SI{1000}{G_0} (\SI{100}{G_0}, \SI{7000}{G_0})\\
		Ini. dust to gas ratio & $\delta_{\rm dg}$ & 0.01423$^{(a)}$\\
		Frag. velocity & $v_\mathrm{frag}$ & \SI{1e3}{cm/s} ($\pm$1 dex)\\
		Dust grain density & $\rho_{\rm s}$ & \SI{1}{g\per\centi\meter\cubed}\\
		Dust monomer size & $a_{\rm min}$ & \SI{e-5}{\centi\meter} ($\pm$1 dex)\\
		\hline
		\multicolumn{3}{l}{\textbf{References.}}\\
		\multicolumn{3}{l}{ $^{(a)}$ \citet{Santos2003} } \B\\
		\hline
	\end{tabular}
	\vspace{1ex}
	
	{\raggedright \textbf{Note.} Results for variations in brackets are presented in Sect. \ref{sec:disk_param}.\par}
\end{table}
An important step is how to initialize the simulations. The starting point of the gas disk is an arbitrary time with an initial surface density profile following Eq. \eqref{eq:initial_surf_dens}. For our calculations, we used the disk parameters listed in Table \ref{tab:modelParams}. However, immediately after initialization the gas disk is in no steady-state. It will evolve on a short timescale until sink terms in Eq. \eqref{eq:disk_evol} are balanced and the disk reaches a quasi-equilibrium. Given the dependency on initial parameters, we do not yet insert the solids and start their evolution during this stage. Instead, we track the outermost cell with $\Sigma_{\rm g}\neq 0$. Once it moves less than \SI{5}{\percent} in the last \SI{1000}{yr} of simulation time, we consider the disk equilibrated. After this stage is reached, we multiply the gas surface density by a constant dust to gas ratio (Table \ref{tab:modelParams}) which results in an initial dust surface density. For the further discussion, we set our time zero to the time at which this happens. The delay of the solid evolution compared to the gas disk as well as the initial solid disk masses can be read off Table \ref{tab:startingPoints}. We note that it is not possible to find an equilibrium for too large values of $\mathcal{F}_{\rm FUV}$ or too low viscosities. This limits the explorable range of parameters and when testing increased values of $\mathcal{F}_{\rm FUV}$, we chose \SI{7000}{G_0} where an equilibrium is still reached.

\section{Results}

\subsection{Model comparison}
\label{sec:model_comparison}
\subsubsection{The magnitude of dust entrainment}
\label{sec:photoevap_results}

\begin{figure}
	\centering
	\includegraphics[width=\linewidth]{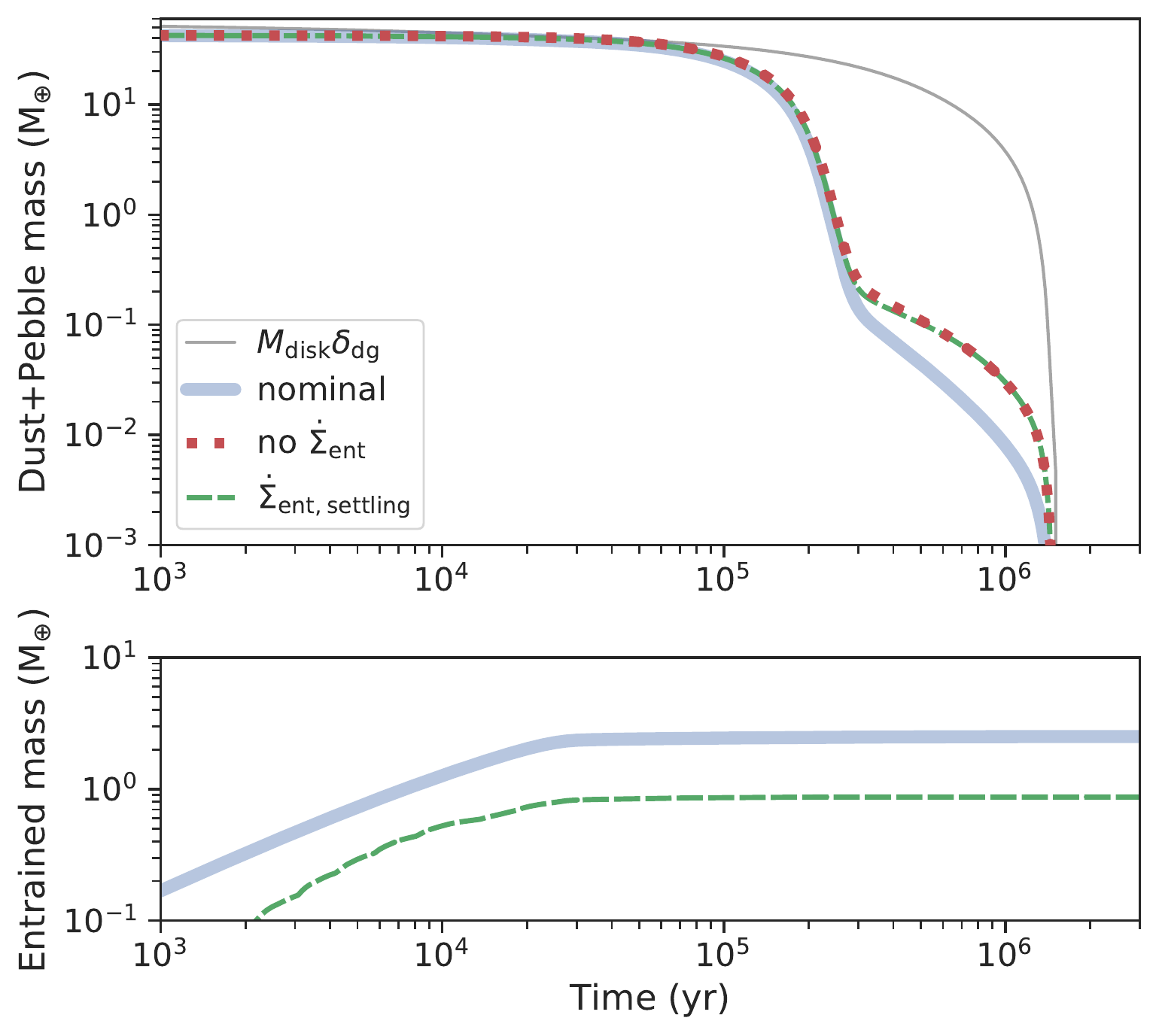}
	\caption{Dust and pebble mass evolution for different dust entrainment models. The top panel shows the mass content at a given time after the start of the solid evolution while the lower panel shows the cumulative mass entrained in evaporative winds. Different calculations are shown for the nominal dust entrainment model (thick, blue), a model without any dust entrainment (red, dotted), and the model with settled dust (dashed, green, see Appendix \ref{app:settled_entrain}).}
	\label{fig:mass_comparison_models}
\end{figure}

\begin{figure}
	\centering
	\includegraphics[width=\linewidth]{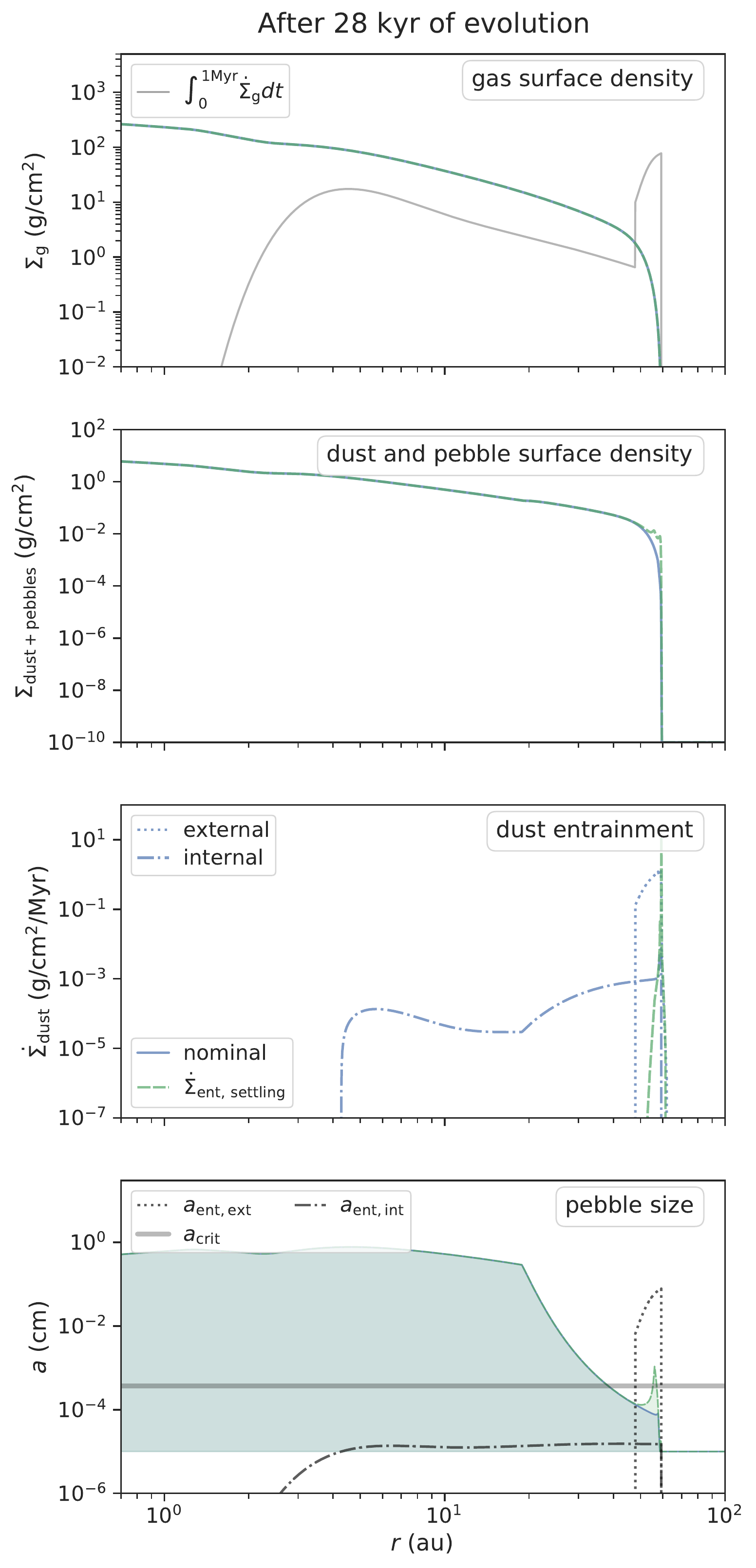}
	\caption{Snapshot of the gas and dust disk after \SI{10}{kyr} of solid evolution for the nominal and the settling model. The two upper panels show the surface density of gas (identical for both cases) and combined dust and pebbles, the third panel the dust and pebble removal rate, and the bottom panel the typical pebble size $a_1$. The shaded region spans from the monomer grain size to $a_1$. In addition to the two different entrainment models, critical entrainment sizes at the midplane (bottom panel) and the integrated gas photoevaporation rate (top panel) are shown as listed in the upper left corners.
	}
	\label{fig:sigma_sigmadot_radial_10kyr}
\end{figure}

\begin{figure}
	\centering
	\includegraphics[width=\linewidth]{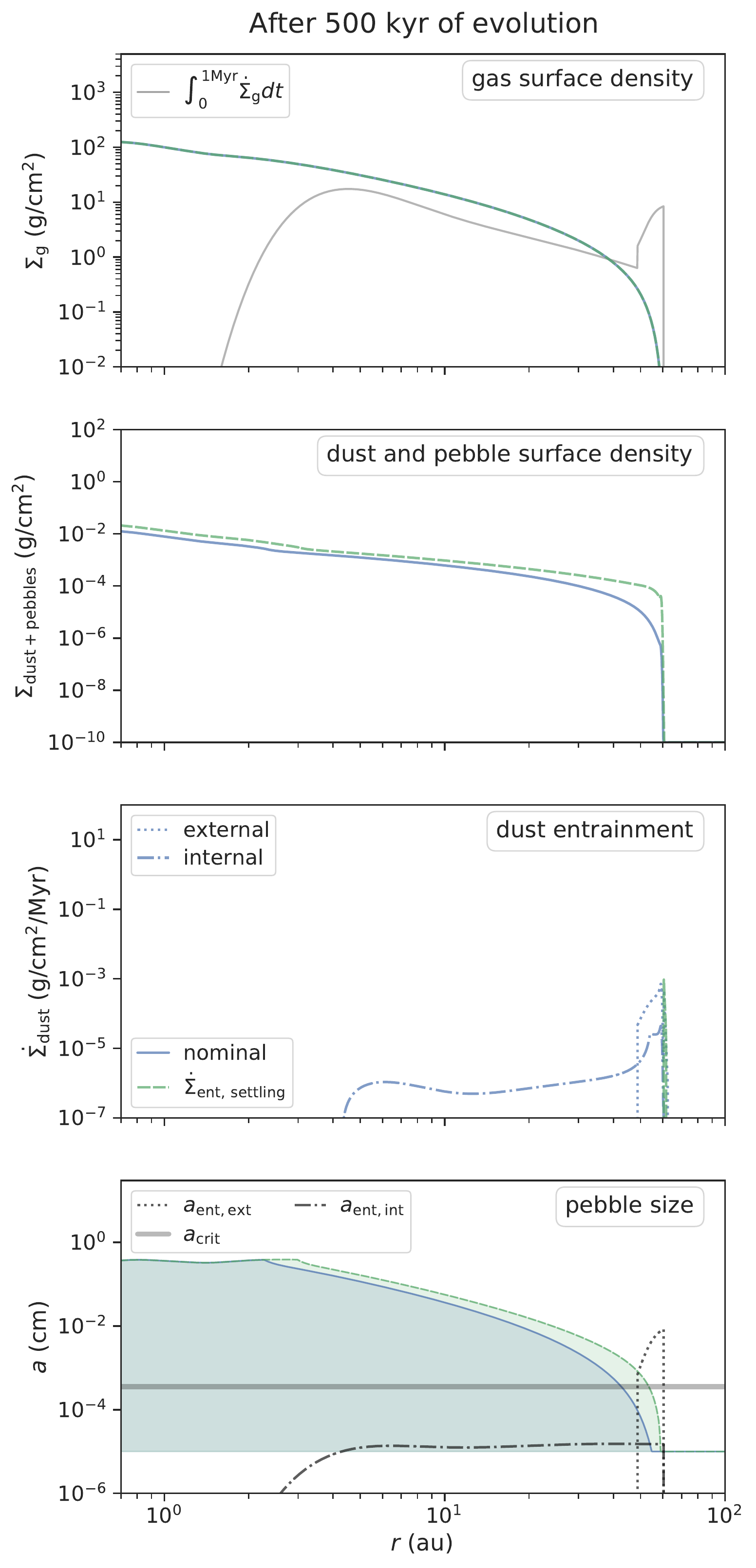}
	\caption{As Fig. \ref{fig:sigma_sigmadot_radial_10kyr}, but after \SI{500}{kyr} of evolution.
	}
	\label{fig:sigma_sigmadot_radial_500kyr}
\end{figure}
For the disk parameters given in Table \ref{tab:modelParams}, we compare here and in the following section the evolution models for dust entrainment and radiation pressure effects described in Sects. \ref{sec:late_stage} and \ref{sec:radiation_pressure_model}, respectively. The chosen nominal values are realistic starting points \citep[e.g.][]{Tobin2020} but arbitrarily chosen for the sake of visualizing the different processes. The nominal external UV field strength $\mathcal{F}_{\rm FUV}$ causing external photoevaporation is on the stronger side and should therefore allow for a comparison of the dust entrainment in the evaporated gas for the different models.

Comparing in Fig. \ref{fig:mass_comparison_models} the red, dotted line for the case without any dust entrainment to the nominal model, we see that dust entrainment can become an important effect in the later stages for the overall mass budget. Several Earth masses of dust can be carried away by the photoevaporative wind. Furthermore, we find that assuming the dust to be settled without the possibility of advection to the base layers will reduce the entrained mass by an order of magnitude.

We further stress that the \SI{\sim 3}{\mearth} which are entrained in winds for the nominal case do not necessarily decrease the disk mass by this amount at a given time. Instead, less mass is accreted onto the star and the disk mass differs by a few tenths of an Earth mass.

In the radial evaporation profiles shown in the third panels of Figs. \ref{fig:sigma_sigmadot_radial_10kyr} and \ref{fig:sigma_sigmadot_radial_500kyr}, we split up the contributions of internally and externally driven photoevaporation for the nominal model (for the settling model only external photoevaporatoin is relevant, see below). The entrainment rates are orders of magnitude larger due to external photoevaporation compared to the internal one. The reason becomes clear from the fourth panel in Fig. \ref{fig:sigma_sigmadot_radial_10kyr} showing the size limits $a_{\rm ent, int}$ and $a_{\rm ent, ext}$ compared to the shaded region spanning over the size distribution of grains in the disk. $a_{\rm ent, ext}$ lies above the largest sizes of dust in the disk meaning the full size distribution can be entrained. Early on, this is because dust did not yet have the time to grow. Later on, the drop in surface density leads to a steepening pressure gradient which in turn lowers the drift limit (Eq. \ref{eq:drift-size}) to retain only small dust (Fig. \ref{fig:sigma_sigmadot_radial_500kyr}). For these reasons the whole size distribution falls below $a_{\rm ent, ext}$ for our nominal disk parameters shown here and the full dust content can be entrained in the flow ($\dot{\Sigma}_{\rm ent} = \delta_{\rm dg}\dot{\Sigma}_{\rm g}$).
	
In contrast, $a_{\rm ent, int}$ barely reaches our chosen monomer grain size of \SI{1e-5}{\centi\meter}. In this case, $f_{\rm ent}$ is small (Eq. \ref{eq:entrained_fraction}) and suppresses entrainment to a large degree. However, it is not zero which will lead to some dust entrainment. At radii smaller than \SI{\sim 4}{\au}, $a_{\rm ent, int}$ falls below the monomer size and no dust is entrained anymore.

We note that in a test without the advection limit of \citet{Booth2021} and using the $a_{\rm ent, ext}$ limit (Eq. \ref{eq:aent_ext}) also for internal photoevaporation instead, we found the entrained mass in internally driven photoevaporative flows to be much lower than in externally driven ones. This is in part due to our choice of parameters (relatively low $L_{\rm X}$ and large $\mathcal{F}_{\rm FUV}$, see Sect. \ref{sec:discussion_internal}) and further facilitated by the much larger area which is covered by external photoevaporation (not well visualized in a radially logarithmic plot).

For the case of a disk with settling and without vertical dust advection (described in Appendix \ref{app:settled_entrain}), we find that dust is entrained in the evaporative flow only where external photoevaporation is acting. At early times, $z_\mathrm{base}$ approaches the midplane in this region. This can only occur if $\Sigma_{\rm g} \Omega_\mathrm{K}$ becomes similar to $\dot{\Sigma}_\mathrm{g}$ (i.e. the mass loss prescription implies mass loss over a dynamical time), which is only the case in the outermost region of the disk where the profile falls off exponentially. For the more evolved disk shown in Fig. \ref{fig:sigma_sigmadot_radial_500kyr}, it is even more rare. Only a single numerical cell has a dust entrainment rate sufficient to appear on the chart. Everywhere else, dust entrainment is negligible. In this settling description, we find that dust settles such that it can be described with a scale heights of a few tenths of the gas scale height for the large particles \citep{Fromang2009} and \num{\sim1}{} gas scale height for the small dust while the base layer of the evaporative flow $z_{\rm base}$ lies at \num{\sim5} gas scale heights in the disk region where internal photoevaporation is effective. This explains the almost full suppression in the settling model case.

Recently, \citet{Franz2022} also assumed that particles remain settled towards the midplane. They used two dimensional ($r,z$) maps of disks undergoing internal XEUV-driven photoevaporation \citep{Picogna2019} and inserted Lagrangian particles which they could track. The entrainment rates found would amount to a few Earth masses of dust which is lost. This shows that in principle, internal evaporation can also lead to significant entrainment of dust under the settling assumption. However, their initial disk size is considerably larger than our disk which is truncated by external evaporation. The $r$-dependent entrainment rates found by \citet{Franz2022} drop to zero within a few tens of astronomical units, consistent with our work. Furthermore, the choice of the assumed lower limit of the size distribution $a_{\rm min}$ influences the results. In our nominal scenario, we used $a_{\rm min} = \SI{0.1}{\micro\meter}$, whereas the distribution of \citet{Franz2022} extends to an order of magnitude smaller grains. We discuss the influence of $a_{\rm min}$ in Sect. \ref{sec:vary_amin}.

Furthermore, the critical entrained sizes $a_{\rm ent, int}$ we derive with our analytic approach for the internal photoevaporation can be compared to more detailed models. \citet{Owen2011a}, \citet{Hutchison2016}, and \citet{Franz2020} found critical sizes ranging from \SIrange{1}{10}{\micro\meter}. Those lie above our limiting size because we consider the advection limit of \citet{Booth2021}. Without this, we would recover similar values. For external photoevaporation, larger grains can be entrained due to the increased evaporation rates based on the FRIED grid and the larger distance to the star. At the outer edge, where there is barely any gas left, the critical size calculated using midplane condition reaches at the early stages of the disk evolution up to \SI{\sim0.1}{\centi\meter}.

We further note that entrainment results would change if we did insert the dust together with the gas instead of waiting for equilibration of viscous spreading and disk truncation due to external photoevaporation (Sect. \ref{sec:setup}). In that case, more of the initially placed dust would be entrained.

Here, we have demonstrated that for a disk extending all the way to the star, internal photoevaporation is not efficient in entraining dust neither under the settling  nor under the advection assumption. However, the picture would change if we increase the X-ray luminosity (see Sect. \ref{sec:vary_LX}) or for a disk with a cavity. Then, as we further discuss in Sect. \ref{sec:caveats_Owen}, radiation can reach the cavity edge and could remove particles at this location \citep{Owen2019}. Our model is in that case incomplete as it considers all the dust to be at least partially shielded from radiation as long as gas is present. Furthermore, with the nominal high turbulent $\alpha$ value and moderate X-ray luminosities $L_\mathrm{X}$ parameters chosen, no inner disk cavity can open.

\begin{figure}[ht]
	\centering
	\includegraphics[width=\linewidth]{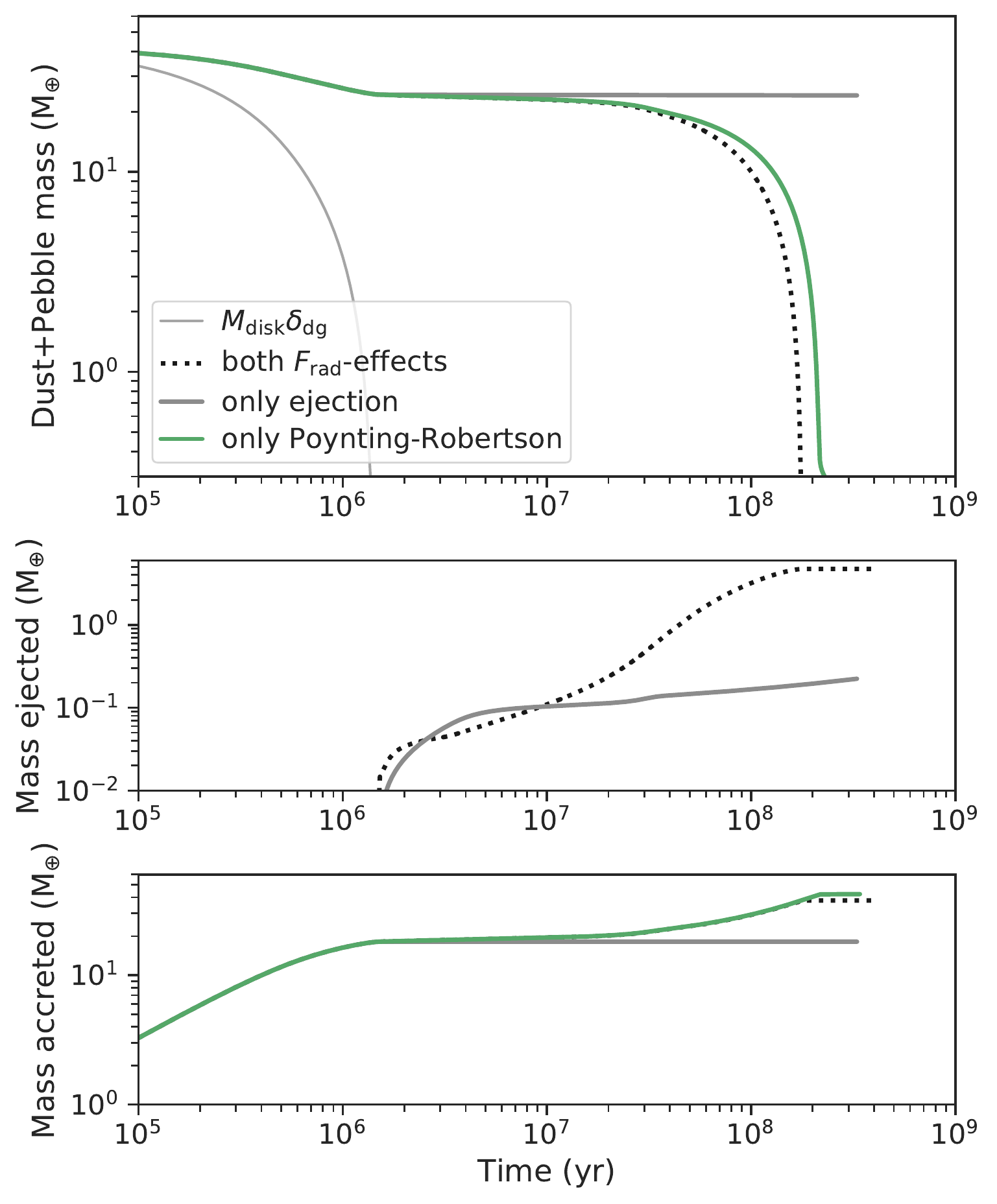}
	\caption{Long-term dust and pebble disk mass evolution (top) as well as dust mass ejected due to radiation pressure (middle) and accreted onto the star (bottom). All models start with a low $v_{\rm frag}$ of \SI{1e-2}{\centi\meter\per\second} and do not include the effect of dust entrainment in photoevaporative winds. The  thick gray line denotes the model without Poynting-Robertson drag but with direct ejection of the smallest grains due to radiation pressure. The inverted assumptions are shown with a green line. The nominal model (dotted, black) includes both effects and is equivalent to the low $v_{\rm frag}$ case shown in the top row of Fig. \ref{fig:mass_comparison_param}.}
	\label{fig:low_alpha_radiated}
\end{figure}

\subsubsection{The importance of Poynting-Robertson drag}
\label{sec:radiation_pressure_results}
In order to study the effects of Poynting-Robertson drag and radiation pressure ejecting dust grains, we change the settings compared to the tests above. Here, we ignore all kinds of entrainment in photoevaporative winds and reduce the fragmentation velocity to \SI{100}{\centi\meter\per\second}. This value is one order of magnitude lower than the nominal setting but still reasonable \citep{Steinpilz2019}. Those changes to the model are made in order to be left with a significant dust and pebble disk once the gas has cleared. In particular, the lowered fragmentation threshold suppresses growth and drift of pebbles. Thus, \SI{24}{\mearth} out of initially \SI{42.5}{\mearth} of small dust is present when radiation from the central star starts to reach dust in our model. In this way, we can test the processes related to radiation pressure, namely Poynting-Robertson drag (Eq. \ref{eq:PoyntingRobertson}) and ejection of grains with sizes $a<a_{\rm crit}$ (Eq. \ref{eq:acrit_rad}) by radiation pressure.

In Fig. \ref{fig:low_alpha_radiated}, the gas disk dispersal is marked by the thin gray line -- showing the gas disk mass multiplied by the assumed dust to gas ratio $\delta_{\rm dg}$ -- dropping to zero at \SI{1.43}{\mega\year}. Subsequently, the disk made of dust and pebbles remains present up to \SI{\sim200}{\mega\year}. During this time, using the nominal model, dust is both ejected by radiation pressure (\SI{5}{\mearth}) and accreted onto the star (another \SI{19}{\mearth} in addition to the \SI{18}{\mearth} which were already accreted during the gas stage). In contrast, if Poynting-Robertson drag is disabled, very little mass is neither ejected nor accreted.

These results demonstrate the interplay of both radiation related effects. Without Poynting-Robertson drag moving dust grains towards the star, very little mass is ejected. This can be attributed in our simple model to the long collisional timescale in the cold and low-density (small $T$ and $\rho_{\rm d}$ in Eq. \ref{eq:tcol}) outer disk. For a massive remnant dust disk, particles generally grow to sizes larger than the critical size everywhere in the disk. Thereafter, only the fraction of the size distribution below $a_{\rm crit}$ can be removed and needs to be replenished by further collisions. If those do not happen frequently enough, only very little mass will be ejected from the system. For the case without Poynting-Robertson drag shown as the gray line in Fig. \ref{fig:low_alpha_radiated}, we find significant ejection of grains out to \SI{\sim5}{\au} only. Another factor is the stellar evolution \citep{Baraffe2015} which leads to a reduction of $L_{\rm star}$ over timescales \SI{\sim1e8}{\year}. Thus, $a_{\rm crit}$ further decreases and a smaller fraction of the dust can be removed at later times.

For these relatively massive dust disks, a similar behavior of the mass evolution is recovered if only Poynting-Robertson drag is included and direct ejection of grains is neglected. However, the amount of accreted mass onto the star will be overestimated because all the mass accretes onto the star instead of a fraction being ejected. This might slightly influence measured stellar metallicities. Nevertheless, to first order, it is more relevant to consider Poynting-Robertson drag compared to direct ejection of grains.

\subsection{Influence of model parameters}
\label{sec:disk_param}
\begin{table}
	\caption{Gas disk equilibration times and initial solid disk masses}
	\label{tab:startingPoints}
	\begin{tabular}{r c c}
		\hline\noalign{\smallskip} 
		Model & Starting time (yr) & Initial solid mass (\SI{}{M_{\oplus}})\\
		\hline\hline\noalign{\smallskip}  
		nominal$^b$ & \SI{18425}{} & 42.5 \\
		high $M_{\rm disk}$ & \SI{15770}{} & 456.8 \\
		low $M_{\rm disk}$ & \SI{19139}{} & 4.2 \\
		low $\alpha$ &  \SI{22892}{} & 43.0 \\
		high $\alpha$ & \SI{13237}{} & 37.5 \\
		low $L_X$ & \SI{18537}{} & 42.6 \\
		high $L_X$ & \SI{18297}{} & 41.6 \\
		low $\beta_\mathrm{g}$ &  \SI{17388}{} & 46.1 \\
		high $\beta_\mathrm{g}$ & \SI{18625}{} & 37.5 \\
		low $r_{\rm out}$ & \SI{13703}{} & 44.8 \\
		high $r_{\rm out}$ & \SI{24283}{} & 21.1 \\
		low $\mathcal{F}_{\rm FUV}$ & \SI{26979}{} & 50.7\\
		high $\mathcal{F}_{\rm FUV}$ & \SI{32834}{} & 30.9 \\
		\hline
		\multicolumn{3}{l}{\textbf{Notes.}}\\
		\multicolumn{3}{p{\linewidth}}{ $^{(b)}$ For higher and lower values of $v_\mathrm{frag}$ and $a_{\rm min}$ the gas evolution is equal to the nominal one; thus, the starting time and initial solid mass remain the same.}\\
		\hline
	\end{tabular}
\end{table}

\begin{figure*}
	\centering
	\includegraphics[width=\linewidth]{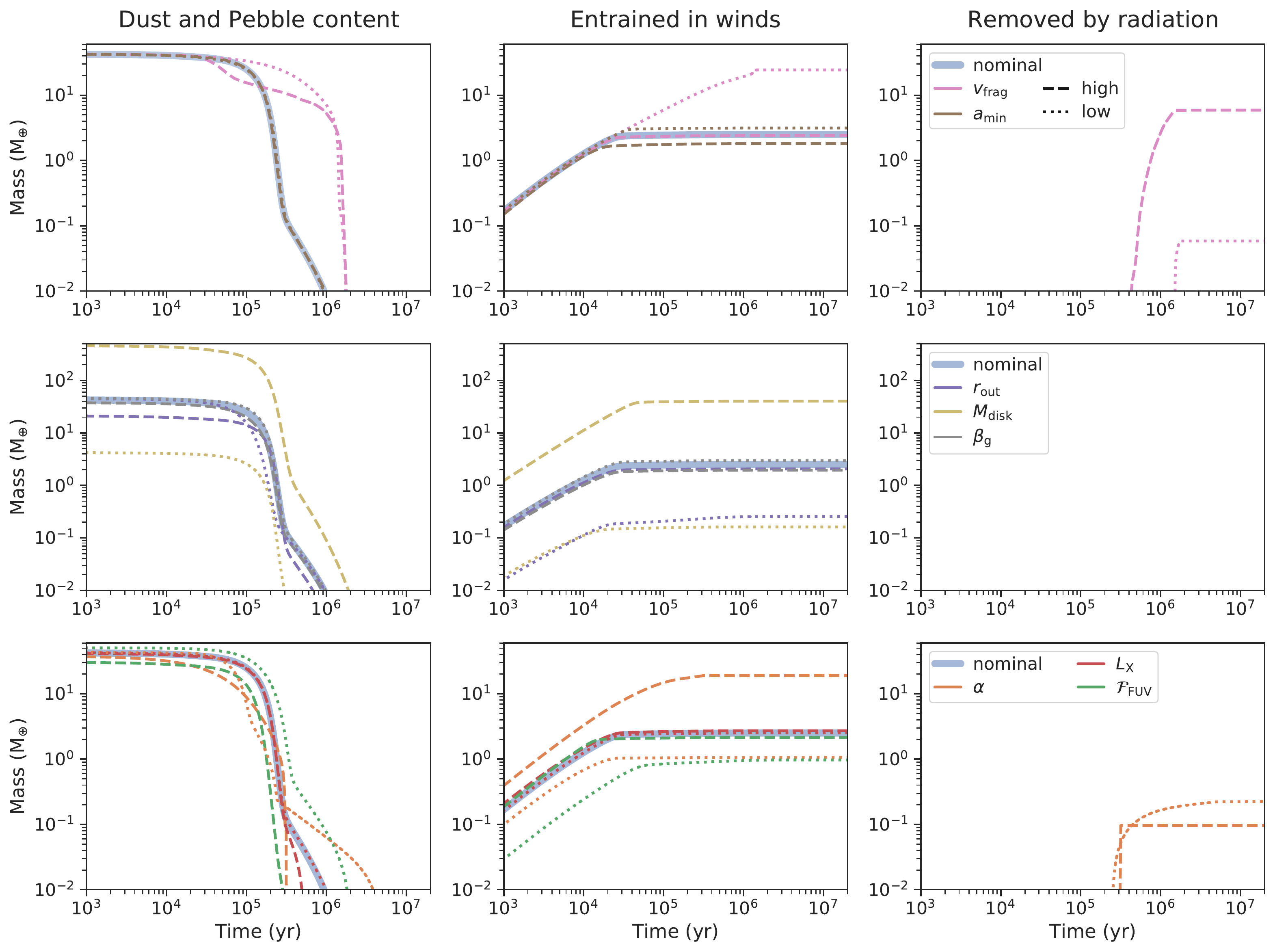}
	\caption{Dust and pebble mass evolution for different parameters. The dashed (dotted) line corresponds to enlarged (lowered) values by the variation indicated in Table \ref{tab:modelParams}. All simulations were done using the nominal dust photoevaporation model and including removal and Poynting-Robertson drag of dust due to radiation pressure. The top row shows the influence of dust-evolution parameters, the middle one results for varied initial disk conditions, and the bottom one shows simulations with different gas-evolution parameters. For several parameters, radiation-removed masses lie below \SI{1e-2}{\mearth} and are therefore not visible.
	 }
	\label{fig:mass_comparison_param}
\end{figure*}

After analyzing the relevance of radiation pressure and dust entrainment in photoevaporative winds for nominal parameters, we explore the dependence on those using a grid of values. We modify the parameters indicated in Table \ref{tab:modelParams}, by generally an order of magnitude to higher or lower values. For $\mathcal{F}_\mathrm{FUV}$, where an increase of an order of magnitude is not possible we chose \SI{7000}{G_0} instead and we chose 0.7 and 1.2 for the lower and higher variations of $\beta_\mathrm{g}$.
The results are presented in Fig. \ref{fig:mass_comparison_param}.

\subsubsection{Initial disk}
\label{sec:vary_mass_profile}
We varied the gas disk mass, exponential cut-off radius as well as the slope of the gas disk and show the results in the middle row of panels in Fig. \ref{fig:mass_comparison_param}. For variations of the profile, the disk mass was kept the same.{Changing the slope or increasing the outer radius di} not significantly influence the results. This is mainly because the gas disk evolves to a very similar state before we start the solid evolution.

For smaller initial disks, less dust is located in the outer regions where it can be removed. This was not fully balanced by the initial \SI{13}{\kilo\year} stage of viscous spreading.

Varying the total disk mass changes considerably the amount of dust which can be entrained. While an order of magnitude change could be expected from the initial solid content, the larger disk with a gas mass of \SI{0.12}{\msol} and a solid content of \SI{456.8}{\mearth} instead of \SI{0.012}{\msol} and \SI{42.5}{\mearth} results in sixteen times more dust entrainment (\SI{40.1}{\mearth} of dust, or \SI{8.8}{\percent} of the initial solid mass). Similar to the results in Sect. \ref{sec:photoevap_results}, dust is mainly removed due to externally induced winds. For a more massive disk, the surface area subject to external photoevaporation is increased because its viscous spreading balances evaporation
at \SI{\sim80}{\au} compared to \SI{60}{\au} for the nominal disk. Therefore, more mass is removed and critical entrainment sizes are larger. Those effects enhance entrainment in a super-linear fashion for more extended disks.

\subsubsection{Photoevaporation parameters}
\label{sec:vary_LX}
Both $L_{\rm X}$ and $\mathcal{F}_{\rm FUV}$ regulate the strength of disk photoevaporation. Increasing those values reduces the gaseous disk lifetime and therefore accelerates the overall evolution. Lowering or increasing external evaporation by the amount listed in Table \ref{tab:modelParams} changes the lifetime of the gas disk by about \SI{1}{Myr}. A similar outcome is found when modifying $L_{\rm X}$, although a small gas disk is present for \SI{\sim4}{Myr} with a low $L_{\rm X}$ of \SI{1e28}{\erg\per\second}.

Using the nominal model, dust is mainly entrained in external photoevaporative winds (see Sect. \ref{sec:photoevap_results}). Thus, the variation of the parameter responsible for the internal photoevaporation, $L_X$, barely changes the amount of entrained dust. However, for variations in the external parameter $\mathcal{F}_\mathrm{FUV}$, significant differences can be seen when lowering the field strength to \SI{100}{G_0}. In this case, the dust mass which is entrained is reduced. Maybe less expected is that an increase in the external UV field does not automatically lead to a larger entrained dust mass. This is because the disk shrinks initially to a smaller size (\SI{\sim35}{au}) which then reduces the area which is subjected to external photoevaporation. As external photoevaporation dominates in all cases studied here, we continue the discussion for reduced $\mathcal{F}_{\mathrm{FUV}}$ and increased $L_X$ in Sect. \ref{sec:discussion_internal}.

\subsubsection{Fragmentation velocity $v_{\rm frag}$}
\label{sec:vary_vfrag}
A change to $v_{\rm frag}$ will affect the local maximum size of the pebbles as $a_1\propto v_{\rm frag}^2$ if fragmentation limits growth (Eq.~\ref{eq:frag-size}) which is the case in the inner disk \citep{Birnstiel2016}. For the same gas disk, we can see in Fig. \ref{fig:mass_comparison_param} that less dust and pebbles are present in the first few \SI{100}{\kilo\year} for a higher $v_{\rm frag}$ (pink dashed line). This is the stage where the front at which pebbles reach large stokes numbers is in the the region of the disk where growth is limited by fragmentation. Therefore, an increased $v_{\rm frag}$ directly translates to larger pebbles which drift faster and pile up at the inner edge. It is worth pointing out, that we chose not to include pebble accretion onto planets \citep{Ormel2010} nor the formation of planetesimals out of the pebble flux \citep[e.g.][]{Lenz2019}. Without those processes and without pressure bumps, pebbles are free to drift to the single pressure maximum close to the inner edge of the disk.

Due to this accumulation, the surface density of dust and pebbles can locally approach and even surpass the surface density of gas and radial drift becomes suppressed (Eq. \ref{eq:radial_drift_w_feedback}, see also \citealt{Nakagawa1986}). Nevertheless, diffusion, Poynting-Robertson drift and ejection of grains in the innermost region where radiation can reach, and (reduced but still present) advection with the gas slowly removes the reservoir of drifted pebbles. This removal of the inner disk is however slower if larger sizes are reached. Thus, in the top left panel of Fig. \ref{fig:mass_comparison_param}, we can see that the dust mass in the nominal case drops below the high-$v_{\rm frag}$ case after a while. We note that in this case with a large accumulation of pebbles in the innermost region, we would expect streaming instabilities \citep{Johansen2007,Klahr2020a,Klahr2020} to occur and potentially trigger the collapse of clumps of dust \citep{Gerbig2020} which is not included in our simulations but will be added in the future.

The optical thickness gradually decreases with time, which allows for a larger region to be radiated as the disk thins out. Thus, the inner dust disk can be removed by radiation pressure thanks to short fragmentation timescales despite the increased particles sizes. We recall that we remove small dust on the longer of the fragmentation or orbital timescale. In the inner disk, where the dust is deposited in this high-$v_{\rm frag}$ scenario, both are short. Furthermore, particles can still be accreted onto the star due to Poynting-Robertson drag. Those effects become important during the later stages of the gas disk evolution and lead to a removal of the dust disk together with the gas.
\\

For low values of $v_{\rm frag}$, which were recently favored by laboratory experiments (\citealp{Steinpilz2019,Gundlach2018,Musiolik2019}; see also \citealp{Pinilla2021}\footnote{We note that \citet{Pinilla2021} propose to modify the vertical turbulence which has to first order the same effect as changing $v_{\rm frag}$.} ) we see less depletion of dust and pebbles in the initial stages. Indeed, for the nominal choice of the rest of the parameters (see Table \ref{tab:modelParams}), a $v_{\rm frag}$ of \SI{100}{\centi\meter\per\second} results in a maximum grain size at any location in the disk below \SI{0.01}{\centi\meter}. Small grains like this essentially follow the gas -- with one important exception: they still settle to regions closer to the midplane. However, with vertical dust advection they are easily transported to the upper layers and therefore still entrained in the photoevaporative wind in our nominal entrainment model.

Due to the small size of the particles for low $v_{\rm frag}$, radial drift is suppressed. Therefore, mass remains in the outer disk for a longer time to be entrained by photoevaporation and entrainment becomes the dominating dust removal effect. The suppression of drift also leaves a larger reservoir of dust after the gas disk has dissipated. Therefore, a light remnant disk of dust (\SI{\sim0.05}{\mearth}) is cleared by radiation pressure in this scenario which is an order of magnitude more than in the nominal case. We note that for the low $v_{\rm frag}$-case but with the settling model instead, we would be left with a much larger remnant almost identical to the test case presented in Sect. \ref{sec:photoevap_results}.

\subsubsection{Dust monomer size $a_{\rm min}$}
\label{sec:vary_amin}
The minimum size of grains is expected to be inherited from the interstellar material with sizes ranging from \SI{0.01}{\micro\meter} to \SI{1}{\micro\meter} \citep{Mathis1977} which are the two limits we explore in addition to the nominal value of \SI{0.1}{\micro\meter}. We find a moderate impact of the initial monomer size on dust entrainment in photoevaporative winds. The entrained mass varies by a factor of \SI{\sim1.5}{} for order of magnitude changes in $a_{\rm min}$. As discussed in Sect. \ref{sec:model_comparison}, the whole size distribution is usually entrained in the externally photoevaporated part of the disk which makes the smallest size irrelevant. However, the magnitude of internal photoevaporation is affected and leads to the aforementioned change. For the high minimum size of \SI{1}{\micro\meter}, no dust is entrained in internally induced photoevaporative winds as this lies above the $a_{\rm ent, int}$ limit of \citet{Booth2021} for the nominal X-ray luminosity (but see also the discussion in Sect. \ref{sec:discussion_internal}).

\subsubsection{Viscous $\alpha$}
\label{sec:vary_alpha}
The interpretation of the effect of the viscous $\alpha$ is more challenging as many aspects depend on it. First, lowering (increasing) $\alpha$ has a similar effect on the dust evolution as an increased (decreased) $v_{\rm frag}$: it changes the maximum size of pebbles due to fragmentation ($a_1\propto \alpha^{-1}$). Similar to the case discussed above, the maximum grain size for high $\alpha$ values remains below a few \SI{100}{\micro\meter}. However, Fig. \ref{fig:mass_comparison_param} shows that for high $\alpha$, the overall evolution is faster which is due to the faster gas disk evolution; the gas disk has disappeared after $\sim$\SI{200}{kyr}. Overall, a very similar amount of dust has been entrained in the high-$\alpha$ case as for low $v_{\rm frag}$, but on a shorter timescale.

We note, that settling would also be influenced for such high values of $\alpha$. Therefore, we found in a test run with the settling model which ignores vertical dust advection (Appendix \ref{app:settled_entrain}) that a similar amount of dust is entrained as for the nominal model. This can be attributed to the vigorous turbulence combined with the small grain size increasing the scale height of dust (Eq. \ref{eq:dtg_z}). Given the current lack of clear mechanisms driving turbulence in disks, it would be possible that $\alpha$ is not constant in time. Therefore, it is important to mention that dust entrainment is already a relevant sink of dust mass if $\alpha$ is on the order of \SI{1e-2}{} for a duration of $\sim$\SI{100}{kyr}. It does not have to remain at this value during the full several Myr of typical disk lifetimes \citep{Haisch2001,Richert2018,Michel2021}. 
\\

Lowering $\alpha$ to \SI{3.16e-4}{} is motivated by recent findings: Observational (e.g. \citealp{Flaherty2017} but see also \citealp{Dullemond2018,Rosotti2020}) and theoretical arguments for low values were put forward in recent years \citep{Bai2013,Klahr2014}. In that case, pebbles can grow similar to the high-$v_{\rm frag}$ scenario. The most considerable fraction of solids grow, drift, and pile up at the disk inner edge and would have been available for planetesimal formation or pebble accretion on the way there.

Again similar to the high-$v_{\rm frag}$ case is the dispersal of the inner dust and pebbles due to radiation pressure and accretion onto the star driven by the Poynting-Robertson effect. The difference is that less dust piles up; thus this mechanism is less pronounced.

We further note that the low value of $\alpha$ reduces the amount of gas that is accreted onto the star and therefore fails to meet the available observational constraints on accretion rates in young clusters \citep[][]{Alcala2014,Alcala2017,Manara2016a,Manara2017a,Manara2020}. Here, we do not yet include magnetized winds \citep{Bai2016} driving accretion onto the star which could resolve this issue and should be explored in the context of dust entrainment in the future.

\section{Discussion}
\label{sec:discussion}

\subsection{Entrainment in internally driven wind}
\label{sec:discussion_internal}
Here, we want to put into perspective our finding that dust is mainly entrained in externally driven photoevaporative winds and not in winds caused by XEUV radiation as described in \citet{Picogna2019,Picogna2021,Ercolano2021}. This conclusion is somewhat warped because our nominal parameters contain a relatively weak X-ray luminosity of \SI{1e29}{\erg\per\second} but a strong FUV field strength of \SI{1000}{G_0} \citep[see e.g.][]{Adams2010}. The former value is more than an order of magnitude lower than the values observed by \citet{Gudel2007} for Solar mass stars. In contrast, FUV field strengths can be much lower than our chosen nominal values if no massive O- or early B-star is present in the vicinity. This is the case for several star forming regions which are observable today \citep[see e.g. the discussion in Sect. 5.5. of][]{Michel2021}. For these reasons, we additionally explore here the case of a disk with negligible external photoevaporation but stronger internal one.

\begin{figure}
	\centering
	\includegraphics[width=\linewidth]{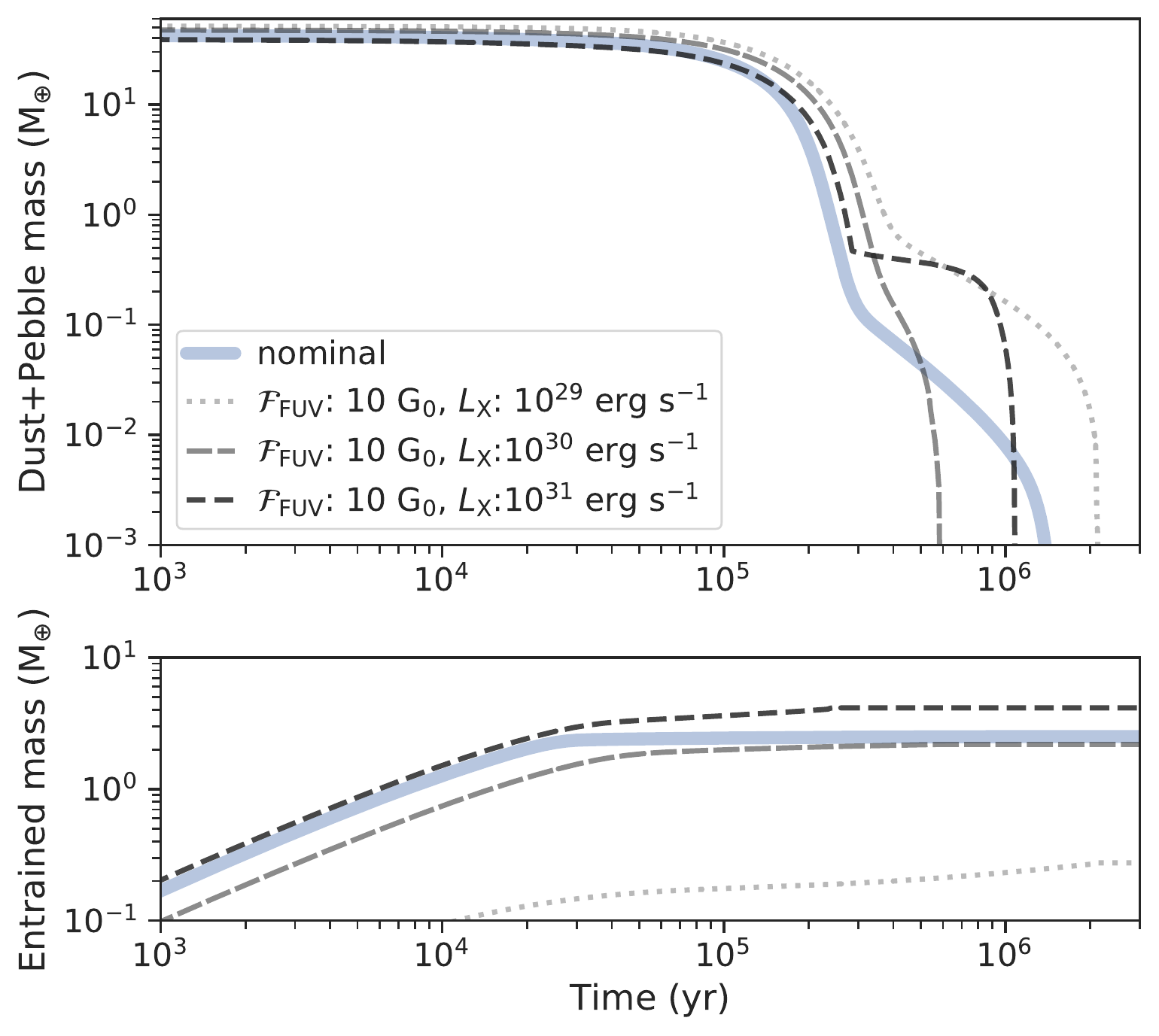}
	\caption{Dust and pebble mass (top) and cumulative entrained mass (bottom) evolution for different X-ray luminosities. $L_{\rm X}$ controls the strength of internally driven photoevaporation in our prescription from \citet{Picogna2021}. Except for the nominal case, we use reduced $\mathcal{F}_{\rm FUV}$.}
	\label{fig:mass_comparison_LX}
\end{figure}

As we show in Fig. \ref{fig:mass_comparison_LX}, it is not at all clear that external evaporation needs to drive a larger dust mass loss compared to internal photoevaporation. For reasonable values of $L_{\rm X}\sim\SI{1e30}{\erg\per\second}$, we find a comparable mass loss as in the mainly externally driven nominal case. An even larger amount can be entrained when the X-ray luminosity is further increased to \SI{1e31}{\erg\per\second}. Interestingly, in this case we obtain a ring feature in the dust at \SI{10}{\au} which survives the gas phase. It emerges due to a bump in the XEUV photoevaporation profile of \citet{Picogna2021} at this location. This dust is then removed by radiation pressure. Such a feature is not present for the lower luminosities.
 
The results shown here highlight that both internal or external photoevaporation can drive significant dust mass losses. Both drivers can dominate for reasonable choices of X-ray luminosities or external field strengths. Depending on the level of extinction of external radiation \citep[see also][]{Cleeves2016} it is possible that internally driven fields dominate also for a more general case than only in regions without massive stars.

\subsection{Caveats}

\subsubsection{Limitations of the two-population approach}
\label{sec:caveats_2pop}
In order to have a model for which the computation time is short enough to use it in population synthesis models, we chose the two-population approach to dust evolution of \citet{Birnstiel2012}. However, there exist some shortcomings compared to a full dust growth and collision model \citep[e.g.][]{Birnstiel2010}. For our application, the most relevant shortcoming is the assumption that gas is always the dominant constituent in the disk. We improved on this and introduced in Sect. \ref{sec:radiation_pressure_model} a regime where Brownian motion and Poynting-Robertson drag replace the collision timescale and the drift velocity. Furthermore, we include the prescriptions of \citet{Nakagawa1986} and \citet{Garate2020} to smoothly reduce radial drift velocities as gas densities reduce and dust starts to dominate and affect the gas.

However, while dust velocities are adjusted consistently, we did not yet include a consistent model for dust sizes. The fragmentation limited maximum size is proportional to $\Sigma_{\rm g}$ and the drift limit depends on the gaseous pressure gradient. Therefore, for fragmentation dominated regions, the size reduces as the gas surface density reduces (keeping the Stokes number constant). Also, in the drift limited case (in the outer region), dust sizes decrease as the gaseous disk shrinks and the pressure gradient steepens towards the outer edge. We stop following this reduction of the representative larger size $a_1$ as soon as gas midplane densities fall below dust midplane densities. This is certainly a threshold where the assumption of gas-dominated evolution fails. However, even before that, it remains to be checked if collisions are frequent enough to grind the pebbles into dust in regions where gas densities are low. Similarly, we do not check if the drift timescale is fast enough to allow for the size reduction which we find towards the outer edge. Since gas evaporation can be relatively quick in regions of low surface densities at a few \SI{}{\au}, the sizes might instead be frozen earlier. A comparison to the results of a full dust growth and collision model (using dustpy, Gárate et al., in prep) will shed light on whether this assumption was warranted.

On a similar note, the implementation of Poynting-Robertson dragged dust is not completely consistent. This is because as particles drift, the two-population approach assumes that they encounter other particles and reach an equilibrium size distribution at the new location. For gas dominated disks this has been verified and is a reasonable assumption \citep{Birnstiel2012}. However, for the case of pure dust disks collisions are more rare and gentle, potentially allowing for the preservation of the size of a drifting body or even growth. Here, we never find prominent radial motion due to Poynting-Robertson drag. Therefore, this is not a major drawback but the size evolution should be revised for studies of debris disks or other massive dust disks without gas.

\subsubsection{Dust feedback on photoevaporation}
We also note that the entrained dust would influence the photoevaporation rate \citep{Facchini2016}, which is not taken into account in our model. While the analytical work of \citet{Facchini2016} sheds some light on these questions, it would be highly useful to run numerical simulations of multi-dimensional disks subject to external photoevaporation including embedded dust particles similar to the works of \citet{Hutchison2016,Franz2020,Franz2022} which address pure internal evaporation. For internal photoevaporation, \citet{Franz2022a} found potentially observable cone or chimney like features which are more pronounced if the dust is considered as settled. Thus, observations in scattered and polarized light will also help to answer those questions.

\subsubsection{\citet{Owen2019} mechanism: Radiation pressure effects in gas-rich disks}
\label{sec:caveats_Owen}
Recently, \citet{Owen2019} proposed a novel mechanism to clear dust from a disk still containing gas with an inner \SI{10}{\au} cavity. Radiation pressure can remove dust above the disk photosphere and therefore clear dust more quickly than gas. The process requires small grains to be reproduced by fragmentation and is generally not efficient for the overall disk \citep{Takeuchi2003}. However, for the case of a pressure trap where the surface density of dust is significantly enhanced, collisional grinding is faster and the dust densities at all heights are enlarged. For disks with large external photoevaporation which are the main focus of this work, the scenario is not relevant. This is because there is no stage where an extended, \SI{10}{\au} cavity opens.

However, for the future development of a model that is applicable for population synthesis where also disks with low $\mathcal{F}_{\rm FUV}$ values need to be explored, the mechanism should be included. As stated by \citet{Owen2019}, the difficulty lies in the coupled problem of the dust evolution influencing the opacity of the disk which in turn is very crucial to determine the photosphere and how much dust lies above it. In general, such a coupling of dust evolution to opacities leads to interesting effects to be explored \citep{Savvidou2020}.

\section{Summary and Conclusions}
\label{sec:conclusion}
In order to better compare models of disk evolution and planet formation to observations, we present an improved description of dust and pebble evolution in the two-population approach that is better applicable to disks which become gas depleted. The model includes dust entrainment in photoevaporative winds under consideration that larger dust is mostly settled to the midplane but dust below a critical size can be transported to the upper layers of the disk. The adopted prescription for internally driven photoevaporation by \citet{Booth2021} is for the first time used in global models while treatment of entrainment in externally driven winds is identical to the approach presented in \citet{Sellek2020a}. Where applicable, the limit of \citet{Booth2021} gives slightly smaller entrainment sizes compared to what was found by \citet{Franz2020,Franz2022,Garate2021a} showing the importance of this effect. Furthermore, we include direct ejection of grains and Poynting-Robertson drag caused by radiation pressure of the central star once the disk becomes optically thin, which will be useful to study the fate of second generation dust \citep{Gerbig2019} in the future.

In this first paper of a series working towards a population synthesis of protoplanetary disks, we vary model assumptions, parameters, and initial conditions to study their effects on the disk mass budget and radial profile. We find:
\begin{itemize}
	\item For internal photoevaporation, the base of the evaporative flow lies an order of magnitude higher than dust scale heights obtained under the assumption that dust settles and no vertical dust advection occurs. Therefore, this layer would be dust-depleted without vertical advection of small grains. Consequently, the ability to advect small grains is typically limiting entrainment rates \citep[in agreement with][]{Booth2021}.
	\item Due to the geometry of externally driven flows, we assume that grains can be entrained from the full vertical extent of the disk. We find that for nominal disk parameters about \SI{5}{\percent} of the solid mass can be lost to externally induced flows.
	\item More dust can be entrained if the disk is more massive (\SI{\sim9}{\percent} for a \SI{0.12}{\msol} disk) or if particles remain smaller due to turbulence or lower fragmentation velocities. For a viscous $\alpha=\SI{3.16e-2}{}$ or a $v_{\rm frag}=\SI{1}{\meter\per\second}$, a total of \SI{\sim50}{\percent} of dust is entrained in the evaporative flow. In this case, grains are not drifting toward the star and become entrained in the photoevaporative flow instead.
	\item For our fiducial parameters, more dust is entrained in externally driven winds. However, if field strengths are reduced as in several star forming regions observable today, entrainment in internally driven winds can also dominate and entrain a similar amount of dust.
	\item Nominally, dust is following the gas and removed as the gas disk disappears. This is only possible thanks to the included vertical transport of small grains. If this is suppressed or if second generation dust forms, we find that a massive debris disk can survive up to \SI{100}{\mega\year}. In this later stage, Poynting-Robertson drag is crucial to move dust towards the star where it can more easily be removed.
\end{itemize}

With this work, we present an important step to pave the way for global models of planet formation to include a more realistic treatment of dust. Detailed emission modeling and comparing to ALMA data will further constrain the nature of the protoplanetary disks in which planets form. Furthermore, when including forming planets and their feedback on the dust disk, we will learn about planet formation as it is happening right now.

\begin{acknowledgements}
	We thank T. Lichtenberg and M. Gárate for fruitful discussion. We further thank the anonymous referee for their insightful comments.
	R.B. acknowledges the financial support from the SNSF under grant P2BEP2\_195285.
	Parts of this work were supported by the DFG Research Unit FOR2544 “Blue Planets around Red Stars”, project no. RE 2694/4-1.
	T.B. acknowledges funding from the European Research Council (ERC) under the European Union’s Horizon 2020 research and innovation programme under grant agreement No 714769 and funding by the Deutsche Forschungsgemeinschaft (DFG, German Research Foundation) under grants 361140270, 325594231, and Germany's Excellence Strategy - EXC-2094 - 390783311.
	The plots shown in this work were generated using \textit{matplotlib} \citep{Hunter2007} and \textit{seaborn} (\href{https://seaborn.pydata.org/index.html}{https://seaborn.pydata.org/index.html}).
\end{acknowledgements}

\bibliography{library_remo,library_add}{}

\appendix
\section{Dust entrainment of settled dust}
\label{app:settled_entrain}
In addition to the dust entrainment model presented in Sect. \ref{sec:late_stage}, we explore the case of a completely settled disk. As discussed in the main text, this is not adopted as the nominal model because the effect of vertical advection of dust grains is not accounted for in this case \citep{Booth2021}. Nevertheless, it is an assumption still under discussion \citep{Hutchison2021,Franz2022}.

We start with the model of \ctS20, but instead of taking $\delta_{\rm dg} = \Sigma_{\rm dust}/\Sigma_{\rm g}$ and midplane values for $q$, we explore the case of a vertically settled distribution of dust without advection. To study this, we calculated a $z$ dependent volume density of dust. To pinpoint the location of the base of the photoevaporative flow, we needed to determine the height where the gas density is (\citealp{Hollenbach1994}, eq. 3.1 or \citealp{Owen2012}, eq. 3)
\begin{equation}
\rho_{\rm g}(z=z_{\rm base}) = \dot{\Sigma}_{\rm g} / (2 c_s)\,,
\label{eq:base_density}
\end{equation}
where $c_s$ is the isothermal sound speed that we used as typical speed of the photoevaporative wind. We note that a more detailed treatment for the wind speed \citep[e.g.][]{Hutchison2016} is not required for our purposes and would vary by less than a factor 4. Furthermore, at large heights, neither the assumption of constant vertical gravity ($z\ll r$) nor that the disk is isothermal in the vertical direction hold. Therefore, a more complete picture could be obtained in future works by accounting for this by numerically calculating the vertical disk temperature profile and setting the base of the photoevaporative flow to the layer where the temperature exceeds the escape temperature \citep{Ercolano2009}.

For a hydrostatic vertical disk, the squared elevation of the base of the flow above the midplane -- measured in gas scale heights -- using the simple approach of Eq. \eqref{eq:base_density} is thus
\begin{equation}
\label{eq:base_of_flow}
z^2_{\rm base}/H_{\rm g}^2 =  \max\left\{ 2 \ln\left( \frac{ 2 \Sigma_{\rm g}c_s}{\sqrt{2\pi} \dot{\Sigma}_{\rm g} H_{\rm g}}\right) , 0\right\}\,,
\end{equation}
where we introduced a lower limit of zero to avoid nonphysical imaginary $z$ values.
We note that this height is only approximate for large elevations but is sufficiently precise for our purposes. To assess the influence of allowing for arbitrarily large $z_{\rm base}$ values, we will further conducted a test where the maximum $z_{\rm base}$ is limited to $2 H_{\rm g}$ which led to some negligible, but non-zero entrainment also in internally driven winds for nominal parameters.

With this value we can move to the dust component of the disk. For a given diffusion coefficient $D$, \citet{Fromang2009} solved the equations of vertical diffusion and settling of dust in steady state. We divide their result \citep[][Eq. 19]{Fromang2009} by the gas density to get
\begin{equation}
\delta_{\rm dg}(z) = \delta_{\rm dg, mid} \exp\left[ - \frac{\mathrm{St}_{\rm mid} \mathrm{Sc}}{\alpha} \left(\exp\left( \frac{z^2}{2 H_{\rm g}^2}\right) -1 \right)\right]\,,
\label{eq:dtg_z}
\end{equation}
where $\mathrm{Sc}$ is the Schmidt number, which we set to unity and therefore neglect terms of $\mathcal{O}\left(\mathrm{St}^2\right)$ in the exponent \citep{Youdin2007,Birnstiel2016}.

To make use of the two-population model, we evaluated expression \eqref{eq:dtg_z} at $z = z_{\rm base}$ for both the large ($\delta_{\rm dg, mid, 1} \approx \delta_{\rm dg, mid} f_m$) and the small ($\delta_{\rm dg, mid, 0} \approx \delta_{\rm dg, mid} (1-f_m)$) particle sizes with midplane Stokes numbers $\rm{St}_{\rm mid, 1}$ and $\rm{St}_{\rm mid, 0}$ to get
\begin{equation}
\delta_\mathrm{dg, 0,1}(z_{\rm base}) = \delta_{\rm dg, mid, 0,1} \exp\left[ - \frac{\mathrm{St}_{\rm mid, 0,1}}{\alpha} \left(\sqrt{\frac{2}{\pi}}\frac{\Omega_\mathrm{K} \Sigma_{\rm g}}{\dot{\Sigma}_{\rm g}} -1 \right) \right]\,.
\end{equation}
Those can then be used to calculate a dust-evolution- and height-dependent slope of the size distribution
\begin{equation}
q(z_{\rm base}) = \frac{\ln\left(\frac{\delta_{\rm dg, 1}}{\delta_{\rm dg, 0}}\right)}{\ln\left(\frac{a_1}{a_0}\right)} + 2\,,
\end{equation}
where the summand of two enters due to changing the base from masses to sizes.

Furthermore, we can use $\delta_{\rm dg, 1}$ and $\delta_{\rm dg, 2}$ in Eq. \eqref{eq:dust_photoevap}. For this last step, we assumed that the two dust populations at the base of the wind are dominated by the $\exp(-\mathrm{St})$ term in Eq. \eqref{eq:dtg_z} and thus we simply summed up the two contributions ($\delta_{\rm dg}(z_{\rm base}) \approx \delta_{\rm dg, 1} + \delta_{\rm dg, 0}$) because $\int_{a_\mathrm{min}}^{\infty} \exp(- a) da = \exp(- a_\mathrm{min})$).

\end{document}